\pdfoutput=1
\documentclass[cernpreprint, texlive=2023, english, texmf, orcidlogo]{atlasdoc}
 
\usepackage{todonotes}
\DeclareBoolOption[true]{LANGEDIT}
\LEcontact{Chip Brock brockr@msu.edu}

\usepackage{atlaspackage}[subcaption]
 
\usepackage{atlasbiblatex}
 
\usepackage{atlasphysics}

\graphicspath{{logos/}{figures/}}
 
\usepackage{atlasbsm}
\usepackage{ANA-SUSY-2019-23-PAPER-defs}

\AtlasTitle{A search for top-squark pair production, in final states containing a top quark, a charm quark and missing transverse momentum, using the 139 fb$^{-1}$ of $pp$ collision data collected by the ATLAS detector}

\AtlasAbstract{
This paper presents a search for top-squark pair production in final states with a top quark, a charm quark and missing transverse momentum. The data were collected with the ATLAS detector during LHC Run 2 and correspond to an integrated luminosity of 139\,fb$^{-1}$ of proton--proton collisions at a centre-of-mass energy of $\sqrt{s} = 13$~$\text{TeV}$. The analysis is motivated by an extended Minimal Supersymmetric Standard Model featuring a non-minimal flavour violation in the second- and third-generation squark sector. The top squark in this model has two possible decay modes, either $\tilde{t}_1\rightarrow c\tilde{\chi}_1^0$ or $\tilde{t}_1\rightarrow t\tilde{\chi}_1^0$, where the $\tilde{\chi}_1^0$ is undetected. The analysis is optimised assuming that both of the decay modes are equally probable, leading to the most likely final state of $tc + E_{T}^{miss}$. Good agreement is found between the Standard Model expectation and the data in the search regions. Exclusion limits at 95\% CL are obtained in the $m(\tilde{t}_1)$ vs $m(\tilde{\chi}_1^0)$ plane and, in addition, limits on the branching ratio of the $\tilde{t}_1\rightarrow t\tilde{\chi}_1^0$ decay as a function of $m(\tilde{t}_1)$ are also produced. Top-squark masses of up to 800~$\text{GeV}$ are excluded for scenarios with light neutralinos, and top-squark masses up to 600~$\text{GeV}$ are excluded in scenarios where the neutralino and the top squark are almost mass degenerate. }

\AtlasRefCode{SUSY-2019-23}
 
\PreprintIdNumber{CERN-EP-2023-262}
 
\AtlasDate{\today}

\AtlasJournal{JHEP}
\AtlasJournalRef{\JHEP 07 (2024) 250}
\AtlasDOI{10.1007/JHEP07(2024)250}

\hypersetup{pdftitle={ATLAS document},pdfauthor={The ATLAS Collaboration}}
 
\begin{document}
 
\maketitle
 
\tableofcontents

\FloatBarrier
 
\section{Introduction}

The Standard Model (SM) of particle physics is one of the most comprehensively tested theories of nature, yielding predictions that agree with a wide range of experimental measurements. However, several aspects of nature remain unexplained by this model despite its success. Amongst them, the experimental mass of the Higgs boson and the possible existence of a non-baryonic component of the universe, called dark matter (DM)~\cite{Goldberg:1983nd, Ellis:1983ew}, pose some of the most important open questions in current particle physics.
 
Supersymmetry (SUSY) \cite{Golfand:1971iw, Volkov:1973ix, Wess:1974tw, Wess:1974jb, Ferrara:1974pu, Salam:1974ig} is one of the most flexible frameworks extending beyond the SM that can provide  answers to the above questions. By introducing a scalar supersymmetric partner for every chiral component of the standard model fermions,\footnote{For a generic fermion $f$, with chiral components $f_L$ and $f_R$, two scalar fields exist (named respectively $\tilde{f}_L$ and $\tilde{f}_R$).} it can mitigate large radiative corrections to the Higgs boson mass. These naturalness arguments favour light states for the supersymmetric partners of the top quark, the \stopL and \stopR. The two top squark states mix to yield mass eigenstates \stopone and \stoptwo, by convention the former being the lightest. Moreover, in $R$-parity-conserving supersymmetric models~\cite{Farrar:1978xj}, supersymmetric partners are produced in pairs, and the lightest supersymmetric particle (LSP) is stable, providing a viable DM candidate. This LSP is generally assumed to be lightest neutralino (\ninoone).
 
\begin{figure}[htpb]
\begin{center}
\includegraphics[width=0.45\textwidth]{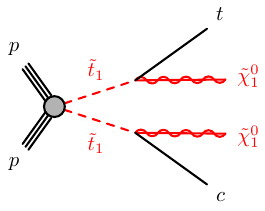}
\end{center}
\caption{Signal diagram for \stopone pair production showing the two possible decay modes of the \stopone. Decays into pairs of top quarks or charm quarks are also taken into account.}
\label{fig:intro_tcmet}
\end{figure}

Searches for the top squark at the LHC experiments have set stringent limits on these particles~\cite{SUSY-2018-12, SUSY-2018-09, SUSY-2018-07, SUSY-2018-08, CMS-SUS-20-002, CMS-SUS-18-004, CMS-SUS-21-003, CMS-SUS-21-004}, imposing constraints on the \stopone mass in the Minimal Supersymmetric Standard Model (MSSM) at the order of 1~\TeV, in scenarios where $R$-parity and flavour are conserved. Extensions of the MSSM can propose scenarios where flavour is not conserved, resulting in looser constraints. In this paper, a non-minimal flavour violation (MFV) extension of the MSSM is considered, as described in Ref.~\cite{Chakraborty:2018rpn}. In this framework, the second-generation and third-generation right-handed squark sectors can mix and the mixing is quantified through the mixing angle ($\theta_{tc}$).
As a consequence of this mixing, the mass eigenstate \stopone is a combination of the gauge eigenstates of the second- and third-generation squarks and can decay into a charm quark and a neutralino in addition to decaying into a top quark and a neutralino, as shown in \Figure\ref{fig:intro_tcmet}. This search focuses on signals containing on-shell top quarks produced in the final state, thus signals with a mass splitting between the top squark and the neutralino (\dmTN) of at least 175~\GeV, and it considers a scan of the branching ratio for $\stopone\to c\ninoone/ t\ninoone$. Only final states with hadronically decaying top quarks are considered in the analysis. The advancement of charm-tagging techniques has opened the exploration of charm particle final states. Those techniques are used in this paper to define a selection predominantly sensitive to events with a top-quark and a charm-quark, allowing a first exploration of this signature at the LHC. Final results are derived by performing model-dependent and model-independent profile likelihood fits, which respectively evaluate the presence of a SUSY signal or a generic one.
 
\FloatBarrier
 
\section{ATLAS detector}

\newcommand{\AtlasCoordFootnote}{
ATLAS uses a right-handed coordinate system with its origin at the nominal interaction point (IP)
in the centre of the detector and the \(z\)-axis along the beam pipe.
The \(x\)-axis points from the IP to the centre of the LHC ring,
and the \(y\)-axis points upwards.
Polar coordinates \((r,\phi)\) are used in the transverse plane,
\(\phi\) being the azimuthal angle around the \(z\)-axis.
The pseudorapidity is defined in terms of the polar angle \(\theta\) as \(\eta = -\ln \tan(\theta/2)\) and is equal to the rapidity
$ y = \frac{1}{2} \ln \left( \frac{E + p_z c}{E - p_z c} \right) $ in the relativistic limit.
Angular distance is measured in units of \(\Delta R \equiv \sqrt{(\Delta y)^{2} + (\Delta\phi)^{2}}\).}
 
The ATLAS detector~\cite{PERF-2007-01} at the LHC covers nearly the entire solid angle around the collision point.\footnote{\AtlasCoordFootnote}
It consists of an inner tracking detector surrounded by a thin superconducting solenoid, electromagnetic and hadronic calorimeters,
and a muon spectrometer incorporating three large superconducting air-core toroidal magnets.
 
The inner-detector system (ID) is immersed in a \qty{2}{\tesla} axial magnetic field
and provides charged-particle tracking in the range \(|\eta| < 2.5\).
The high-granularity silicon pixel detector covers the vertex region and typically provides four measurements per track,
the first hit generally being in the insertable B-layer (IBL) installed before Run~2~\cite{ATLAS-TDR-19,PIX-2018-001}.
It is followed by the SemiConductor Tracker (SCT), which usually provides eight measurements per track.
These silicon detectors are complemented by the transition radiation tracker (TRT),
which enables radially extended track reconstruction up to \(|\eta| = 2.0\).
The TRT also provides electron identification information
based on the fraction of hits (typically 30 in total) above a higher energy-deposit threshold corresponding to transition radiation.
 
The calorimeter system covers the pseudorapidity range \(|\eta| < 4.9\).
Within the region \(|\eta|< 3.2\), electromagnetic calorimetry is provided by barrel and
endcap high-granularity lead/liquid-argon (LAr) calorimeters,
with an additional thin LAr presampler covering \(|\eta| < 1.8\)
to correct for energy loss in material upstream of the calorimeters.
Hadronic calorimetry is provided by the steel/scintillator-tile calorimeter,
segmented into three barrel structures within \(|\eta| < 1.7\), and two copper/LAr hadronic endcap calorimeters.
The solid angle coverage is completed with forward copper/LAr and tungsten/LAr calorimeter modules
optimised for electromagnetic and hadronic energy measurements respectively.
 
The muon spectrometer (MS) comprises separate trigger and
high-precision tracking chambers measuring the deflection of muons in a magnetic field generated by the superconducting air-core toroidal magnets.
The field integral of the toroids ranges between \num{2.0} and \qty{6.0}{\tesla\metre}
across most of the detector.
Three layers of precision chambers, each consisting of layers of monitored drift tubes, cover the region \(|\eta| < 2.7\),
complemented by cathode-strip chambers in the forward region, where the background is highest.
The muon trigger system covers the range \(|\eta| < 2.4\) with resistive-plate chambers in the barrel, and thin-gap chambers in the endcap regions.
 
The luminosity is measured mainly by the LUCID--2~\cite{LUCID2} detector that records Cherenkov light produced in the quartz windows of photomultipliers located close to the beampipe.
 
Events are selected by the first-level trigger system implemented in custom hardware,
followed by selections made by algorithms implemented in software in the high-level trigger~\cite{TRIG-2016-01}.
The first-level trigger accepts events from the \qty{40}{\MHz} bunch crossings at a rate below \qty{100}{\kHz},
which the high-level trigger further reduces in order to record complete events to disk at about \qty{1}{\kHz}.
 
A software suite~\cite{ATL-SOFT-PUB-2021-001} is used in data simulation, in the reconstruction
and analysis of real and simulated data, in detector operations, and in the trigger and data acquisition
systems of the experiment.


\FloatBarrier
 
\section{Data and simulated event samples}
\label{sec:dataMC}

 
The proton--proton collision data analysed in this paper were collected between 2015 and 2018 at a centre-of-mass energy of 13~\TeV with a 25~ns proton bunch crossing interval.
Multiple $pp$ interactions occur per bunch crossing (pile-up), with a measured average of approximately 34 interactions.
Application of beam, detector, and data-quality criteria~\cite{DAPR-2018-01} result in a total integrated luminosity of 139~fb$^{-1}$. The uncertainty in the combined 2015--2018 integrated luminosity is 1.7\%~\cite{ATLAS-CONF-2019-021}, obtained using the LUCID-2 detector~\cite{LUCID2} for the primary luminosity measurements and cross-checked by a suite of other systems.
 
This analysis searches for signatures with a significant missing transverse momentum ($\vec{p}_{\mathrm{T}}^{\mathrm{miss}}$). This $\vec{p}_{\mathrm{T}}^{\mathrm{miss}}$ is calculated as the negative sum of all reconstructed objects and a ``soft-term'' composed of all tracks associated with the primary vertex but not matched to any reconstructed object. The magnitude of this \ptmiss\ is the missing transverse energy, denoted as $\met$. Due to the presence of two \ninoone\ which don't interact with the detector in the final state considered in this analysis, events are required to satisfy a $\met$ trigger~\cite{TRIG-2016-01,TRIG-2019-01} and to present an offline reconstructed $\met$ exceeding 250~\GeV~\cite{TRIG-2019-01} in order to ensure the full efficiency of the trigger. To aid with the estimate of some of the SM background processes, events are also selected using single-lepton triggers~\cite{TRIG-2018-01,TRIG-2018-05} with corresponding offline thresholds above 27~\GeV used to ensure the lepton triggers are also fully efficient.
 
Samples of Monte Carlo (MC) simulated events are used to model the SUSY signal and background processes in the analysis. The SUSY signal models were generated with \MGNLO[2.8.1]~\cite{Alwall:2014hca} at leading order (LO) in QCD using the \NNPDF[3.0nlo]~\cite{Ball:2014uwa} set of parton distribution functions (PDFs) and interfaced with \PYTHIA[8.244]~\cite{Sjostrand:2014zea} using the A14 set of tuned parameters (`tune')~\cite{ATL-PHYS-PUB-2014-021} with \NNPDF[2.3lo] for the parton showering and hadronisation, and with \EVTGEN[1.7.0]~\cite{Lange:2001uf} for the modelling of heavy flavour hadron decays. The samples were normalised using the cross-section calculations at next-to-next-to-leading order (NNLO) in the strong coupling constant, adding the resummation of soft gluon emission at next-to-next-to-leading-logarithmic (NNLL) accuracy~\cite{Beenakker:1997ut,Beenakker:2010nq,Beenakker:2011fu}. The signal samples were generated with both possible decays $\stopone\rightarrow t\ninoone$ and $\stopone\rightarrow c\ninoone$ occurring with equal probability. This allows a branching-ratio (BR) scan to be performed by applying a reweighting procedure based upon the number of generated signal events of each decay-type.
 
The SM backgrounds considered in this analysis are: \Zjets\ production; \Wjets\ production; \ttbar\ production; single-top-quark production; \ttbar\ production in association with electroweak bosons ($\ttbar+V$); multi-top and top rare processes ($tWZ$, $tZ$); and multi-boson production ($VV$). The events were simulated using different MC generator programs depending on the process. Details of the generators, PDF set and tune used for each process are listed in Table~\ref{tab:mc_info}.
 
For all the generated samples, the response of the detector was modelled with the full ATLAS detector simulation~\cite{SOFT-2010-01} based on \GEANT~\cite{Agostinelli:2002hh}.  All simulated events were overlaid with multiple $pp$ collisions simulated with \PYTHIA[8.186] using the A3 tune~\cite{ATL-PHYS-PUB-2016-017} and the \NNPDF[2.3LO] PDF set~\cite{Ball:2012cx}. The MC samples were generated with variable levels of pile-up in three campaigns which were reweighted to match the actual distribution of the mean number of interactions observed in data in 2015--2018.
 
\begin{table}[htbp]
\scriptsize
\centering
\caption{Summary of the simulated background samples. `ME' stands for matrix-element, `PS' for parton shower and `UE' for underlying event.}
\vspace{3mm}
\begin{tabular}{llllll}
\toprule
Process & ME event generator and order   & ME PDF & PS and  & UE tune & Cross-section   \\
&                                       &              & hadronisation &         & calculation   \\
\hline
$V$+jets $(V=W/Z)$ & \SHERPA~2.2.1~\cite{Gleisberg:2008ta}, NLO & \NNPDF[3.0nnlo]~\cite{Ball:2014uwa} & \SHERPA       & Default & NNLO~\cite{Catani:2009sm}    \\
Multi-boson            & \SHERPA~2.2.1 or 2.2.2~\cite{Bothmann:2019yzt}, NLO            & \NNPDF[3.0nnlo] & \SHERPA       & Default & NLO \\
$\ttbar$           & \POWHEGBOX~\cite{Alioli:2010xd}, NLO       & \NNPDF[3.0nlo]   & \PYTHIA[8.230]  & A14     & NNLO+NNLL~\cite{Cacciari:2011hy,Czakon:2011xx} \\
Single-top         & \POWHEGBOX, NLO                           & \NNPDF[3.0nlo]   & \PYTHIA[8.230]  & A14     & NNLO+NNLL~\cite{Kidonakis:2011wy, Kidonakis:2010ux, Kidonakis:2010tc}      \\
$\ttbar+V $        & \AMCatNLO[2.3.3], NLO                         & \NNPDF[3.0nlo]     & \PYTHIA[8.210]  & A14     & NLO     \\
$tWZ$, $tZ$              &\AMCatNLO[2.3.3], NLO                         & \NNPDF[3.0nlo]   & \PYTHIA[8.230]  & A14     & NLO        \\
$ttt$, $t\bar{t}t\bar{t}$ & \AMCatNLO[2.3.3], NLO & \NNPDF[3.1nlo] & \PYTHIA[8.230]  & A14 & NLO \\
\bottomrule
\end{tabular}
\label{tab:mc_info}
\end{table}


\FloatBarrier
 
\section{Event reconstruction}
\label{sec:objects}

 
The analysis uses standard ATLAS reconstruction techniques with the object selections used to define small-$R$ jets, large-$R$ jets, leptons ($e, \mu$), top- and $b$-tagged jets reported in Table~\ref{tab:objects}. A loose set of requirements on the properties of the candidate objects is used to define ``baseline'' objects. The baseline requirements for each physics object are shown in the second column of Table~\ref{tab:objects}. The physics objects used to define selections and calculate kinematic variables are required to satisfy tighter selections, referred to as the ``signal'' requirements, and are presented in the third column of Table~\ref{tab:objects}. Baseline objects are used to estimate the $\met$, previously mentioned in Section~\ref{sec:dataMC}. A quality criterion for the matching of topological cell clusters~\cite{PERF-2014-07} in the electromagnetic calorimeter to electrons is also imposed in events containing electrons with $|\eta| \in [1.37,1.52]$ in data recorded during 2015 and 2016.
 
After the baseline objects are defined, an overlap removal procedure is applied to prevent double-counting of tracks and energy depositions associated with overlapping jets, electrons and muons. The procedure applies the following actions to the event. First, baseline electrons are discarded if they share a track in the inner detector with a baseline muon. Next, any jet within a distance $\Delta R=\sqrt{(\Delta y)^2+(\Delta\phi)^2}=0.2$ of a baseline electron is discarded and the electron is retained, if the jet is \textit{not} identified as a $b$-tagged jet, otherwise the jet is kept instead. Similarly, any jet satisfying $N_\text{trk} < 3$ (where $N_\text{trk}$ refers to the number of tracks with $\pT>500~\MeV$ that are associated with the jet) within $\Delta R < 0.2$ of a baseline muon is discarded and the muon is retained. Finally, baseline electrons or muons within a distance $\Delta R < 0.4$ of any remaining jet are discarded.
 
For charm tagging, a non-standard, analysis-specific $c$-tagging algorithm, denoted DL1r$_c$, was optimised. It is based on the DL1r algorithm used for $b$-tagging~\cite{FTAG-2019-07} that provides three probabilities of a jet to be coming from the hadronisation of a $b$-quark ($b$-jet), a $c$-quark ($c$-jet) or a lighter quark/gluon ($l$-jet). These three probabilities are combined in a similar manner as in the $b$-tagging algorithm, but with tuned parameters optimised to identify jets containing $c$-hadrons. To avoid ambiguities in identifying a jet as both a $c$- or $b$-jet, priority is given to $b$-tagged jets. Namely the two tagging algorithms, DL1r and DL1r$_c$, are run in sequence with the $b$-tagging algorithm taking precedence, and if a jet is classified as a $b$-jet, it is no longer considered as an input to the $c$-tagging algorithm. This technique is referred to as \textit{$c$-tagging with $b$-veto} technique and is very helpful to avoid a large rate of $b$-jets misidentified as $c$-jets. The chosen working point for the $b$-tagging algorithm is the 77\% working point, which corresponds to an identification efficiency of 77\% for $b$-jets and a misidentification rate of 20\% for $c$-jets and 0.9\% for light flavour jets. The working point selected for the $c$-tagging algorithm (including the effect of the $b$-veto) corresponds to a 20\% $c$-jet efficiency, with rejection factors of 29 for $b$-jets and 57 for light-jets, evaluated on simulated \ttbar{} events. Similarities in the decay chain of hadronically decaying $\tau$-leptons and c-hadrons lead to a significant misidentification efficiency of $\tau$-leptons as charm jets of the order of 15\%. Due to this, dedicated kinematic variables are defined at event level to reject events containing hadronic $\tau$-leptons, as described in Section~\ref{sec:analysis}. The DNN top tagger~\cite{JETM-2018-03} is used to identify large-$R$ jets which arise from top quark decays. The 80\% efficiency working point is used and it is valid for top-tagged jets with masses in the 40\textendash600~\GeV range and $p_{\mathrm{T}}$ between 350 and 2500~\GeV. Large-$R$ jets outside these validity ranges are not identified as arising from a top quark.
 
Scale factors are applied to account for differences between data and simulation for the trigger, reconstruction, identification, and isolation efficiencies. In the case of the charm tagging algorithm used in this analysis, dedicated scale factors and uncertainties were derived using the same techniques as used for $b$-tagging algorithms~\cite{FTAG-2019-07}. The resulting scale factors correct for differences between data and simulation in its $c$-jet efficiency and in its $b$-jet and light-jet misidentification rate. In the $p_{\mathrm{T}}$ range between 20 and 250~\GeV, the $c$-jet correction factors are found to be mostly compatible with unity, with systematic uncertainties ranging from 17\% at low $p_{\mathrm{T}}$ ($ < 65$ \GeV) to a few \% for higher-$p_{\mathrm{T}}$ jets. The $b$-jet correction and light-jet factors are also found to be compatible with unity, with uncertainties ranging from 7\% to 5\% between 20 and 250~\GeV for $b$-jets and around 13\% for all light-jets. For jets with $p_{\mathrm{T}} > 250$~\GeV, the correction is assumed constant and an additional systematic uncertainty due to the extrapolation is derived using MC simulation, with the uncertainty increasing up to 30\% for charm and bottom jets with \pt{} of 3~\TeV.
 
\begin{table}[htbp]
\footnotesize
\centering
\vspace{3mm}
\caption{Overview of the baseline and signal physics object definitions. The impact parameter along the beam direction and the significance of the transverse impact parameter are denoted by $z_0 \sin\theta$ and $|d_0/\sigma_{d_0}|$, respectively. Similarly, \pt{}, $\eta$, and $m$ are the transverse momentum, pseudorapidity and mass, respectively, for each physics object. WP is the considered working point for each jet tagger, corresponding to the listed selection efficiency.}
\label{tab:objects}
\begin{tabular}{lcc}
\toprule
Physics object &  Baseline requirements   & Additional signal requirements   \\
\hline
Electrons     & Loose likelihood-based selection~\cite{EGAM-2018-01} & Tight likelihood-based selection~\cite{EGAM-2018-01} \\
& $|z_0 \sin\theta|$~<~0.5~mm & \\
& $|\eta|$~<~2.47 & \\
& \pt{}~>~4.5~GeV & \pt{}~>~10~GeV \\
&             & Isolation `Loose' ($\pt > 200$~\GeV `HighPtCaloOnly')~\cite{EGAM-2018-01} \\
&             &  $|d_0/\sigma_{d_0}|$~<~5 \\
\hline
Muons         & Medium identification~\cite{MUON-2018-03} &  \\
& $|\eta|$~<~2.7 & \\
& \pt{}~>~4~GeV & \pt{}~>~10~GeV \\
&             & Isolation `Loose\_VarRad'~\cite{MUON-2018-03} \\
&             &  $|d_0/\sigma_{d_0}|$~<~3 \\
\hline
Small-$R$ jets     & Particle-flow anti-$k_t$ $R$=0.4~\cite{Cacciari:2008gp,Fastjet,PERF-2014-07,PERF-2015-09} & Jet vertex tagger > 0.5 and $|\eta|<2.4$~\cite{PERF-2014-03} \\
& \pt{}~>~20~\GeV & or \pt{}~>~60~\GeV \\
& $|\eta|$~<~2.8 & \\
\hline
Large-$R$ jets     & LCTopo trimmed $R$=1.0~\cite{JETM-2018-02} &  \\
& \pt{}$ \geq 200 $~\GeV & \\
& $|\eta|$~<~2.0 &  \\
\hline
$\met$        &  & Tight WP~\cite{PERF-2016-07}  \\
\hline
Top-tagged jets     &  & DNN top tagger~\cite{JETM-2018-03}  \\
&  & 80\% efficiency WP \\
&  & \pt{}~$\in [350,2500]$\,\GeV  \\
&  & $m$ $\in [40,600]$\,\GeV \\
\hline
$b$-tagged jets     & & DL1r tagger~\cite{FTAG-2019-07}  \\
& & 77\% efficiency WP  \\
\hline
$c$-tagged jets     & & DL1r$_c$ tagger  \\
& & 20\% efficiency WP \\
\bottomrule
\end{tabular}
\end{table}


\FloatBarrier
 
\section{Analysis strategy}
\label{sec:analysis}

The kinematic properties of the SUSY signal under consideration are heavily dependent upon the mass splitting $\Delta m(\stopone, \ninoone)$. Generally, the parameter space can be split into three main regions: the ``bulk'' region, with large $\Delta m(\stopone, \ninoone)$; the ``intermediate'' region where the top squark and neutralino are relatively close in mass; and, finally, the ``compressed'' region, where the mass splitting is such that the top quark from the $\stopone$ decay is produced just on-shell ($\Delta m(\stopone, \ninoone) \approx m_{t}$).
 
The general analysis strategy is to define Signal Regions (SRs) which target different regions of the SUSY signal phase-space by placing selections on particle multiplicities and kinematic variables. Control Regions (CRs) are then defined with negligible signal contamination and enriched in the main backgrounds present in the SRs. These CRs are orthogonal to the SRs. A likelihood fit is performed where normalisation factors for the main backgrounds are calculated in the CRs and extrapolated to estimate the contributions of these backgrounds in the SRs. Finally, Validation Regions (VRs) are defined, again, non-overlapping with the SRs and the CRs, to investigate the modelling of the backgrounds and confirm the normalisation factor can be extrapolated to the SRs.
 
Many kinematic variables are employed in the definition of the SRs to isolate the SUSY signals and to reject events arising from the SM background. In addition to selections on the number of $b$-tagged jets (\nbjets), $c$-tagged jets (\ncjets), large-$R$ jets compatible with arising from a top quark ($N_{\mathrm{tops}}^{\mathrm{DNN}}$), $\met$, and selections on the transverse momentum of the jets ($p_{\mathrm{T}(j/b/c)}$), more complex variables are used, and are described below.
 
\begin{itemize}
\item $\Delta\phi(j_{1-4}, \met)_\mathrm{min}$, $\Delta\phi(j_{1-3}, \met)_\mathrm{min}$: \\ The minimum difference in azimuthal angle between any of the leading four ($j_{1-4}$) or three ($j_{1-3}$) jets and the $\met$. In regions with zero lepton this variable removes the contribution from multi-jet processes which arise from jet mismeasurement leading to ``fake'' $\met$ in the final state.
\item $\mtcmin$, $\mtbmin$:
\\ The minimum value of the transverse mass calculated between any of the $c$-tagged ($b$-tagged) jets and the missing transverse momentum vector. The transverse mass ($m_{\mathrm{T}}$) is defined as:
\begin{center}
$m_{\mathrm{T}} = \sqrt{2p_{\mathrm{T}}(c/b) E_{\mathrm{T}}^{\mathrm{miss}}(1-\cos \Delta\phi(c/b,E_{\mathrm{T}}^{\mathrm{miss}}))}$.
\end{center}
In \ttbar events, these variables present kinematic end-points at truth-level around the mass of the top quark and the \Wboson-boson, which is exploited to reduce the \ttbar background.  
\item $\mtcmax$, $\mtbmax$:
\\ The maximum value of the transverse mass when calculated between any of the $c$-tagged ($b$-tagged) jets and the missing transverse momentum vector. These variables are used to enhance the signal against the \ttbar background, as the \ttbar background tends to have lower values than the signal.
\item $\mtjclose$:
\\ The transverse mass calculated between the jet closest in azimuthal angle to the missing transverse momentum vector and the missing transverse momentum vector itself. This variable is used in conjunction with the $\Delta\phi$ variables to remove the contributions from multijet processes with ``fake'' $\met$.
\item $\met$ Sig:
\\ The global $\met$ significance, calculated including the parameterisations of all objects, defined as:
\begin{center}
$\met$ Sig $= \sqrt{\frac{|\vec{p}_\mathrm{T}^\mathrm{miss}|^2}{\sigma_\mathrm{L}^2(1 - \rho_{\mathrm{LT}}^2)}}$.
\end{center}
Here $\sigma_{\mathrm{L}}$ is the summed momentum resolution of all the objects entering the $\met$ calculation in the direction parallel to the $\ptmiss$ direction (longitudinal direction), while $\rho_{\mathrm{LT}}$ is the correlation factor between this longitudinal resolution and the total momentum resolution in the transverse direction to $\ptmiss$. This variable is used to discriminate between events where the $\met$ arises from invisible particles in the final state and events where the $\met$ arises from poorly measured particles \cite{ATLAS-CONF-2018-038}.
\item $\mttwobinc$:
\\ The stransverse mass variable \cite{Lester:1999tx}, a generalisation of the transverse mass for the case in which two semi-invisibly decaying particles are pair-produced, defined as:
\begin{center}
$m_{\mathrm{T2}}(\vec{p}_{\mathrm{T}}^{1}, \vec{p}_{\mathrm{T}}^{2},\vec{p}_{\mathrm{T}}^{\mathrm{miss}}) =
\underset{\vec{q}_\mathrm{T}^1 + \vec{q}_\mathrm{T}^2 = \vec{p}_{\mathrm{T}}^\mathrm{miss} }{\min} \{  \max[ m_{\mathrm{T}}(\vec{p}_{\mathrm{T}}^{1}, \vec{q}_{\mathrm{T}}^{1} ), m_{\mathrm{T}}(\vec{p}_{\mathrm{T}}^{2}, \vec{q}_{\mathrm{T}}^{2} )] \}. $
\end{center}
This generalization assumes that two visible particles, each reconstructed as one of the objects defined in Section~\ref{sec:objects}, and two invisible particles, reconstructed as $\met$, are produced. A minimization is then performed on the transverse mass of the combinations between the two visible particles and all the possible transverse momenta configurations of the two invisible particles, denoted as $\vec{q}_{\mathrm{T}}^{1}$ and $\vec{q}_{\mathrm{T}}^{2}$,  consistent with the observed $\met$. In this analysis, the two visible objects in the decay, with transverse momentum denoted as $\vec{p}_{\mathrm{T}}^{1}$ and $\vec{p}_{\mathrm{T}}^{2}$, are: a large-$R$ jet in the event containing a $b$-tagged jet ($\Delta\phi(\mathrm{j^{b}_{R=1.0},b}) < 1.0$); and a $c$-tagged small-$R$ jet. If, in a given event, there is more than one large-$R$ jet containing a $b$-tagged jet, or more than one $c$-tagged jet, the large-$R$ jet and $c$-jet with the highest $\vec{p}_{\mathrm{T}}$ are used in the calculation. No top-tagging requirement is imposed on the  $b-$tagged large-$R$ jet. This allows the use of this variable in the definition of all regions across the analysis, regardless if those regions require events with a top-tagged jet or not. For the high-mass \stopone signal this variable is expected to extend to large values, whereas for the \ttbar background this variable should dramatically drop-off at the top quark mass.
\item $\meff$:
\\ The effective mass, the scalar sum of the transverse momenta of all jets in the event and the $\met$, defined as:
\begin{center}
$\meff = \sum_{i = 0}^{\njets} \pt(j_{i}) + \met.$
\end{center}
Generally, for the signal scenarios considered, the $\meff$ extends to high values, compared to the SM backgrounds which generally have lower values of $\meff$.
\end{itemize}
 
Four orthogonal sets of SRs were designed to target the different areas of parameter space, with SRA targeting the bulk region, SRB and SRC targeting the intermediate region, and SRD targeting the compressed region. Due to general similarities between the kinematics of the bulk and intermediate regions, SRA, SRB and SRC have generally similar selections and primarily differ based upon the usage of the \mttwobinc variable and the number of top-tagged jets present in the event. Due to the similarity between the SRA, SRB, and SRC regions, common CRs are used to predict the main SM backgrounds, and common VRs are used to validate the background modelling. The kinematic properties of the SRD region are significantly different, with a selection requiring the presence of initial-state radiation (ISR) to boost the sparticle system, providing enough $\met$ to satisfy the trigger selection. All SRs require that events contain zero leptons, at least one $b$-tagged jet, at least one $c$-tagged jet, and high $\met$.
 
The full selections used to define the A-, B-, and C-type regions are summarized in Table \ref{tab:AB_Selections}. Commonalities among these regions include specific selections based on large values of $\met$, $\met$ significance, and \mtcmin between the $c$-tagged jet and the $\met$. SRA is kept orthogonal to the SRB region using the number of top-tagged large-$R$ jets (\ntops), with SRA requiring at least one top-tagged jet whereas SRB requires exactly zero top-tagged jets. The SRA and SRC regions are orthogonal due to the selection on \mttwobinc. The SRB and SRC regions are also orthogonal due to the selection on \ntops, with SRB requiring zero top-tagged jets whereas SRC requires at least one. 
 
In the parameter space targeted by the SRA region, very large values of \mttwobinc are expected. To enhance sensitivity to these signals when performing the model dependent interpretation, a multi-bin fit is performed, and two bins of \mttwobinc are used: $[450,575]$ and $\geq 575$~\GeV. For the intermediate region of parameter space, lower values of \mttwobinc are expected, so, instead, sensitivity can be enhanced by performing a multi-bin fit in the \mtjclose variable. The SRB region is split into three orthogonal regions in the \mtjclose variable: $[100,150]$, $[150,400]$ and $\geq 450$~\GeV. SRC uses a similar strategy, however, with a finer granularity of \mtjclose bins: $[100,150]$, $[150,300]$, $[300,500]$, and $\geq 500$~\GeV. When performing the model-independent fit, the same lower bounds are used for the SRA, SRB, and SRC regions as in the model-dependent fit, however, the upper bounds on the bins are removed, allowing for more general regions than those used for the model-dependent interpretation.
 
The largest background contribution in SRA, SRB, and SRC originates from \Zjets events, followed by single-top quark (`single-top') or \Wjets\ events. Due to the stringent requirement in $\met$ significance and $\Delta \phi(j_{1-4}, \met)_\mathrm{min}$, the multijet background is negligible.  Common CRs are used to estimate the main backgrounds for both of the SRA and SRB regions, with a single VR used to validate the background estimate strategy. CRs requiring two leptons (2L) are used to extract a normalisation factor for the \Zjets\ background. In these regions the leptons are subtracted from $\vec{p}_{\mathrm{T}}^{\mathrm{miss}}$, and are used as a proxy for neutrinos mimicking the $Z \rightarrow \nu\nu$ decay. This ``lepton-corrected'' $\met$ is denoted by \METprime\ and in the 2L regions is used when calculating all kinematic variables. The CRZAC region is used to normalise the \Zjets\ process in both of the SRA and SRC regions (as both of the regions require one top-tagged large-$R$ jet). The CRZB region is used to estimate the \Zjets\ background in the SRB region, where the presence of a top-tagged large-$R$ jet is explicitly vetoed. A CR requiring exactly one lepton (1L), denoted CRstAC, is used to estimate the single-top contribution in the SRA and SRC regions by specifically requiring the presence of one top-tagged large-$R$ jet in the event. Due to the subdominant contribution of other potential SM backgrounds, no further CRs are defined, and rather the contributions from these processes are estimated directly from simulation. A single 0L VR, denoted VRZABC, is used to validate the modelling of the \Zjets\ background in all three SRs. This 0L selection is inclusive relative to the number of top-tagged jets in the event, and hence can be used to validate the modelling across all SRA, SRB, and SRC regions.
 
\begin{table}[!htb]
\footnotesize
\centering
\caption{Analysis selections for the A-, B-, and C-type regions associated with the SRs targeting the bulk and intermediate signal mass scenarios respectively. Selections denoted by an \textbf{*} are split into multiple bins (for the SR considered) when the model-dependent fit is performed and the selection shown in the table is the lower-bound on the selection used for the given region. `SFOS' indicates that the selected leptons are required to have the same flavour and  opposite-sign electric charges, such that they are compatible with the decay of a \Zboson boson. More information about the fit strategy is provided in the text.}
\begin{tabular}{|c||c|c|c|c|c|c|c|}
\hline
\textbf{Variable} & \textbf{SRA} & \textbf{SRB} & \textbf{SRC} & \textbf{VRZABC} & \textbf{CRstAC}  & \textbf{CRZB} & \textbf{CRZAC} \\
\hline
\hline
Trigger & \multicolumn{5}{c|}{$\met$ Trigger} & \multicolumn{2}{c|}{1L Trigger} \\
\hline
\thead{Baseline \& \\ signal leptons} & \multicolumn{4}{c|}{= 0} & \thead{= 1 \\ (\pt $\geq$ 30 \GeV)} & \multicolumn{2}{c|}{ \thead{= 2 SFOS \\ (\pt $\geq$ 30 \GeV )}} \\
\hline
\thead{$\met$  [$\GeV$]}   & $\geq$ 250 & $\geq$ 300  & $\geq$ 250 & $\geq$ 250 & $\geq$ 250  & \multicolumn{2}{c|}{$\leq$ 150}  \\
\hline
\thead{\METprime[$\GeV$]} & \multicolumn{5}{c|}{--}  & \multicolumn{2}{c|}{$\geq$ 250} \\
\hline
(\nbjets, \ncjets)  & \multicolumn{7}{c|}{($\geq$ 1,$\geq$ 1) (Leading jet is either $b$-tagged or $c$-tagged)}\\
\hline
\njets  & $\geq$ 3 & $\geq$ 5  & $\geq$ 3 & $[3,8]$ & $\geq$ 3  & $\geq$ 5 & $\geq$ 3  \\
\hline
\thead{\ptj{1}  [$\GeV$]} & \multicolumn{5}{c|}{$\geq$ 50} & \multicolumn{2}{c|}{$\geq$ 20}  \\
\hline
$\Delta\phi(j_{1-4},$ $\met$$)_\mathrm{min}$ & \multicolumn{4}{c|}{$\geq$ 0.4}  & \multicolumn{3}{c|}{--} \\
\hline
$m_{\ell\ell}$ [$\GeV$] & \multicolumn{5}{c|}{--} & \multicolumn{2}{c|}{$[76,106]$} \\
\hline
\thead{\mtcmin \\ {[$\GeV$]}} & \multicolumn{4}{c|}{$\geq$ 200} & $\geq$ 300 & $\geq$ 150 & --  \\
\hline
\thead{\mtcminprime \\ {[$\GeV$]}} & \multicolumn{5}{c|}{--} & \multicolumn{2}{c|}{$\geq$ 150} \\
\hline
\thead{$N_{\mathrm{tops}}^{\mathrm{DNN}}$ \\ $(R = 1.0)$} & $\geq$ 1 & = 0 & $\geq$ 1 & -- & $\geq$ 1  & = 0 & $\geq$ 1  \\
\hline
\thead{\ptb{1}  [$\GeV$]} & $\geq$ 20 & $\geq$ 20 & $\geq$ 100 & $\geq$ 20 & $\geq$ 20 & \multicolumn{2}{c|}{$\geq$ 20}  \\
\hline
\thead{\ptc{1}   [$\GeV$]} & $\geq$ 20 & $\geq$ 100 & $\geq$ 100 & $\leq 200$ & $\geq$ 20 & \multicolumn{2}{c|}{$\geq$ 20} \\
\hline
\thead{\ptj{2}  [$\GeV$]} & $\geq$ 20 & $\geq$ 100 & $\geq$ 20 & $\geq 20$  & $\geq$ 20  & \multicolumn{2}{c|}{$\geq$ 20}  \\
\hline
\thead{\ptj{4}  [$\GeV$]} & -- & $\geq$ 50 & -- & -- & --  & \multicolumn{2}{c|}{--}  \\
\hline
\thead{\mtcmax \\ {[$\GeV$]}} & -- & $\geq$ 400  &  -- & $\leq$ 400 & -- & $\geq$ 400 & --  \\
\hline
\thead{\mtbmax \\ {[$\GeV$]}} & -- & $[200,700]$ &  -- & $\geq$ 200 & -- & $\geq$ 200 & -- \\
\hline
\thead{\mtbmin \\ {[$\GeV$]}} & \multicolumn{4}{c|}{$\geq$ 200} & $\geq$ 300  & $\geq$ 200 & --  \\
\hline
\thead{\mtjclose \\ {[$\GeV$]}} & $\geq$ 100 & \multicolumn{2}{c|}{$\geq$ 100 \textbf{*}} & $\geq$ 150 & $\geq$ 200 & \multicolumn{2}{c|}{--} \\
\hline
$\met$ Sig & $\geq$ 18 & $\geq$ 10 & $\geq$ 17 & $[15,17]$ & $[12,22]$  & $\geq$ 10 & $\geq$ 17\\
\hline
\thead{\mttwobinc \\ {[$\GeV$]}} & $\geq$ 450 \textbf{*} & $\geq$ 150 & $[200,450]$ & -- &  $\geq$ 200 & \multicolumn{2}{c|}{--} \\
\hline
 
\end{tabular}
\label{tab:AB_Selections}
\end{table}
\clearpage
 
Due to the small mass difference between the \stop and \ninoone targeted, the kinematics in the SRD region are dramatically different from the other SRs. An ISR-like selection, requiring additional jet activity, is used to ``boost'' the sparticle system providing the large $\met$ to satisfy the trigger selection. The full selections used to define all D-type regions are summarized in Table \ref{tab:C_Selections}.

Generally, the selections for the SRD region follow a standard ISR-like selection with a high-\pt{} leading-jet which is not $b$- or $c$-tagged. Due to the similarity of the signal with the main SM backgrounds of \ttbar and \Vjets, a dedicated multi-class neural network (NN) was developed to isolate the signal against these two main backgrounds.
 
The NN~\cite{WITTEN2017417} is defined using the \textsc{PyTorch} library~\cite{NEURIPS2019_9015} to identify signal events which contain the mixed $tc+\met$ decay, $\ttbar$, and {\ensuremath{\ensuremath{V}\text{+\,jets}}}. To avoid issues due to the finite size of the training dataset, this NN is trained using events which satisfy a looser 0L ISR-like selection (which is a baseline for the SRD definition) and uses a total of 44 low-level variables\footnote{The $p_{\mathrm{T}}, \eta, \Delta\phi(j,\met)$ and flavour of the 6 jets of the event with the highest-$p_{\mathrm{T}}$; the $p_{\mathrm{T}}, \eta, \Delta\phi(j,\met)$ of the 2 highest-$p_{\mathrm{T}}$ $b$-tagged and $c$-tagged jets of the event; the jet, $b$-jet and $c$-jet multiplicities; and the $\met$.} to produce three output scores according to the likelihood for a given event to be signal-like, $\ttbar$-like or \Vjets-like. Binary scores were also considered to separate signal and background events, but they were found to be suboptimal compared to the considered three output scores. A dedicated hyperparameter optimisation is performed with a ``leaky ReLU'' activation function~\cite{Maas2013RectifierNI} and the model is optimized by minimising a cross-entropy loss function. The key variables found to be useful by the NN are $\met$, the $b$-jet multiplicity, and the $\pt$ of the two leading jets. A ``leaky ReLU'' activation function is also applied to the final layer of the NN leading to an output which is a set of scores per event for the three categories mentioned above. The signal score (``NN signal score'') is later used to define SRD. The \Vjets (``\Vjets score'') score is implemented in the \Wjets CR defined below, to enhance the \Wjets contribution. As defining a region enhanced in \ttbar is comparatively simple in the parameter space targeted, the output corresponding to \ttbar events is unused.
 
In the parameter space targeted by SRD, large values of \meff are expected. When considering the model-dependent interpretation, a 2D binning in both \meff and \mtjclose is employed to further enhance sensitivity to the signal. The binning in \meff is as follows: SRD750, $[750,1000]$; SRD1000, $[1000,1250]$; SRD1250, $[1250,1500]$; SRD1500, $[1500,1750]$; SRD1750, $[1750,2000]$; and SRD2000, $\geq 2000$~\GeV. The four lowest \meff bins are further subdivided using the \mtjclose variable: SRD750 is split into four bins $[0,100]$, $[100,200]$, $[200,300]$, and $\geq 300$~\GeV; SRD1000 is split into three bins $[0,100]$, $[100,200]$, $\geq 200$~\GeV; and both SRD1250 and SRD1500 are split into two bins of $[0,100]$ and $\geq 100$~\GeV. In a similar manner to the SRA and SRB regions, when performing the model-independent fit, the lower bound of the \meff selections are used, with the upper bounds removed. For the SRD750 and SRD1000 regions the \mtjclose selection is increased to $\geq 200$~\GeV, whereas for the remaining SRD regions, it is removed.
 
The multijet background is also negligible in the SRD region, its main backgrounds being \ttbar, {\ensuremath{\Wboson\text{+\,jets}}}, and {\ensuremath{\Zboson\text{+\,jets}}}. Three dedicated CRs are defined to estimate the backgrounds from these processes. The CRZZ region is a 2L region where (as in the CRZAC and CRZB regions) the leptons are subtracted from the $\met$ calculation and mimic neutrinos from the $Z$ boson decay, and the lepton-corrected \METprime is used in all calculations. To constrain \Wjets and \ttbar, single-lepton CRs are used (CRWD and CRttD respectively), which are orthogonal due to the \nbjets selections used. In these 1L regions, the lepton is added to the jet collection to mimic the scenario where a hadronic-$\tau$ in the zero-lepton selection is misreconstructed as a jet. Additionally, to ensure a region pure in the \Wjets process, the output \Vjets score of the NN classifier is used. A common 0L VR, denoted VRD, is defined requiring the NN score in the range $[0.0,0.5]$. It is thus orthogonal to the SR, for which the NN score is required to be $\geq 0.75$,  and is then used to validate the extrapolation of the three main backgrounds.
 
As mentioned in Section~\ref{sec:dataMC}, the \ttbar\ background is modelled at NLO using \POWHEGBOX~\cite{Alioli:2010xd}. A mismodelling of the top quark \pt\ distribution was observed for these simulated samples~\cite{TOPQ-2020-02,TOPQ-2018-26}. The  top quark \pt\ is highly correlated with the \meff\ distribution used in the D-regions,  thus CRttD and VRD regions are also binned in \meff in order to respectively correct and validate the \ttbar\ background prediction in the \meff-binned SRD. For CRttD, three bins in \meff are considered: CRttD750, $[750,1000]$\GeV; CRttD1000, $[1000,1250]$~\GeV; and CRttD1250, $\geq 1250$~\GeV. The lack of statistical precision at high-\meff prevents applying a finer binning at high-\meff. This is compensated by a finer binning in the VRD region: VRD750, $[750,1000]$; VRD1000, $[1000,1250]$; VRD1250, $[1250,1500]$; VRD1500, $[1500,1750]$ and VRD1750 $\geq 1750$~\GeV. This latter bin also includes a $\met < 600$~\GeV requirement to minimize a possible signal contamination.
 
\begin{table}[!htb]
\footnotesize
\centering
\caption{Analysis selections for the D-type regions associated with the SR which targets compressed signal scenarios. Selections denoted by \textbf{*} are split into multiple bins when the fitting procedure is performed and the selection shown in the table is the lower-bound on the selection used for the given region. More information about the fit strategy is provided in the text. }
\begin{tabular}{|c||c||c|c|c|c|}
\hline
\textbf{Variable} & \textbf{SRD} & \textbf{CRttD} & \textbf{CRWD}  & \textbf{CRZD} & \textbf{VRD} \\
\hline
\hline
Trigger & \multicolumn{3}{c|}{$\MET$ Trigger} & 1L Trigger & $\MET$ Trigger\\
\hline
\thead{Number of baseline \& \\ signal leptons} &  = 0 & \multicolumn{2}{c|}{\thead{ = 1 \\ (\pt $\geq$ 30 \GeV)} } & \thead{= 2 (\pt $\geq$ 30 \GeV)} & = 0 \\
\hline
\thead{$\met$  [$\GeV$]}   & \multicolumn{3}{c|}{$\geq$ 250} & $\leq$ 100 & $\geq$ 250  \\
\hline
\thead{\METprime[$\GeV$]}& \multicolumn{3}{c|}{--}  & $\geq$ 250 & -- \\
\hline
\njets  & \multicolumn{5}{c|}{$\geq$ 3  (Leading jet not $b$- or $c$-tagged)}  \\
\hline
\ncjets  & \multicolumn{5}{c|}{$\geq$ 1}  \\
\hline
\nbjets  & $\geq$ 1 & $\geq$ 2 & = 1 & $\geq$ 1 & $\geq$ 1 \\
\hline
\thead{\mtcmin \\ {[$\GeV$]} }& \multicolumn{2}{c|}{$\geq$ 100} &  $\geq$ 150 & $\geq$ 100 & $\geq$ 100\\ \hline
\ptj{1} [$\GeV$] & \multicolumn{5}{c|}{$\geq$ 100}  \\
\hline
\ptj{4} [$\GeV$] & \multicolumn{5}{c|}{$\geq$ 30}  \\
\hline
$\met$ Sig & \multicolumn{5}{c|}{$\geq$ 6}  \\
\hline
$\Delta\phi(j_{1-3},$ $\met$$)_\mathrm{min}$ & \multicolumn{3}{c|}{$\geq$ 0.3} & -- & $\geq$ 0.3 \\
\hline
$\Delta\phi(j_{1-3},$ \METprime$)_\mathrm{min}$ & \multicolumn{3}{c|}{--} & $\geq$ 0.3 & -- \\
\hline
$m_{\mathrm{T}}$ [$\GeV$] & -- & $\geq$ 30  & $[30,120]$ & -- & -- \\ \hline
NN \Vjets score & \multicolumn{2}{c|}{--} & $\geq 0$ & -- & -- \\ \hline
\dRjj & \multicolumn{5}{c|}{$\geq$ 1.0} \\ \hline
\dRbl & \multicolumn{2}{c|}{--} & $\geq$ 1.8 & -- & -- \\ \hline
$m_{\ell\ell}$ [$\GeV$] & \multicolumn{3}{c|}{--} & $[76,106]$ & -- \\  \hline
NN signal score & $\geq$ 0.75 & \multicolumn{3}{c|}{$\geq$ 0.0} & $[0,0.5]$ \\ \hline
\meff[$\GeV$]& $\geq 750$ \textbf{*} & $\geq 750$ \textbf{*} & \multicolumn{2}{c|}{--} & $\geq 750$ \textbf{*} \\ \hline
\end{tabular}
\label{tab:C_Selections}
\end{table}
\def\arraystretch{1.}


\FloatBarrier
 
\section{Systematic uncertainties}

This analysis considers several sources of uncertainty, of both
experimental and theoretical nature, that affect the prediction of the
SM background and the SUSY signal in all channels.

The jet energy scale and resolution uncertainties (for both the small-$R$ and large-$R$ jets) are derived as a function of the jet $\pt$, $\eta$, and flavour, using a combination of data and simulated events, as detailed in Refs.~\cite{JETM-2018-05}. These uncertainties also take into account the different pile-up conditions during the four years of data-taking.
 
Uncertainties in the correction factors for the $b$- and $c$-tagging identification efficiencies are applied to the simulated event samples. The corrections are extracted from dedicated flavour-enriched samples in data~\cite{FTAG-2018-01,FTAG-2020-08,FTAG-2019-02}. An additional term is included to extrapolate the measured uncertainties to the high-$\pt$ region of interest. This term is calculated from simulated events by considering variations on the quantities affecting the $b$-tagging performance, such as the impact parameter resolution, percentage of poorly measured tracks, description of the detector material, and track multiplicity per jet. The dominant effect on the uncertainty when extrapolating to high-$\pt$ is related to the different tagging efficiency when smearing the track impact parameters based on the resolution measured in data and simulation.
 
Uncertainties connected with the lepton reconstruction and identification are included in the fit, and they are found to have a negligible impact. All uncertainties in the final-state object reconstruction are propagated to the reconstruction of the $\met$, including an additional term taking into account uncertainties in the scale and resolution of the soft term.
 
The uncertainties related to the modelling of the SM background processes using MC simulation are taken into account. The modelling uncertainties are assumed to be correlated between different SRs but are uncorrelated between the different processes considered. Uncertainties related to the modelling of the \ttbar and single-top processes arise due to the choice of hard-scattering generator and the matching scheme, and these uncertainties are evaluated by comparing the nominal samples with an alternative generator (MadGraph5\_aMC@NLO). The uncertainty due to the choice of PS and hadronisation model is calculated by comparing the nominal sample with a sample produced with Powheg interfaced to \Herwig~7~\cite{Bahr:2008pv,Bellm:2015jjp}. Variations of the renormalisation and factorisation scales, the initial- and final-state radiation parameters and PDF sets are also considered~\cite{ATL-PHYS-PUB-2018-009}. Specifically for the single-top process, a systematic uncertainty corresponding to the interference term between single-top and \ttbar events (at NLO) is applied by comparing the nominal sample, generated with the diagram removal (DR) scheme, with a dedicated sample generated with the diagram subtraction (DS) scheme~\cite{Frixione:2008yi}. The modelling uncertainties related to the \Zjets and \Wjets processes are evaluated using the 7-point renormalisation and factorisation scale variations~\cite{Bothmann:2016nao}, varying these scales by factors of 0.5 and 2. The matrix element matching and the resummation scales are also varied by 0.5 and 2. A conservative 100\% uncertainty is applied to \Zjets events containing both a true $b$- and $c$-quark in the final state. For rare backgrounds that are not normalised in any region (multi-boson, $t\bar{t}V$ and $t+X$), a conservative 30\% uncertainty is applied, covering the difference between the theory prediction for these subprocesses and their measured cross-section by ATLAS plus one standard deviation~\cite{TOPQ-2018-08, TOPQ-2018-01, STDM-2018-34,TOPQ-2016-11, STDM-2017-19, STDM-2018-30}. Similarly, an uncertainty of 5\% is applied on \ttbar and \Wjets processes in regions A, B and C, an uncertainty of 30\% is applied to single-top processes in the D-regions, and an uncertainty of 10\% is applied to \ttbar$Z$ processes in all regions as these backgrounds are not normalized in any region. For the SUSY signal scenarios, systematic uncertainties are also calculated by varying the factorisation and renormalisation scales, the ISR parameters, and the choice of PDF\@. The maximum uncertainty for any signal mass scenario is found to be 20\%.
 
\begin{figure}[htpb]
\begin{center}
\includegraphics[width=1.0\textwidth]{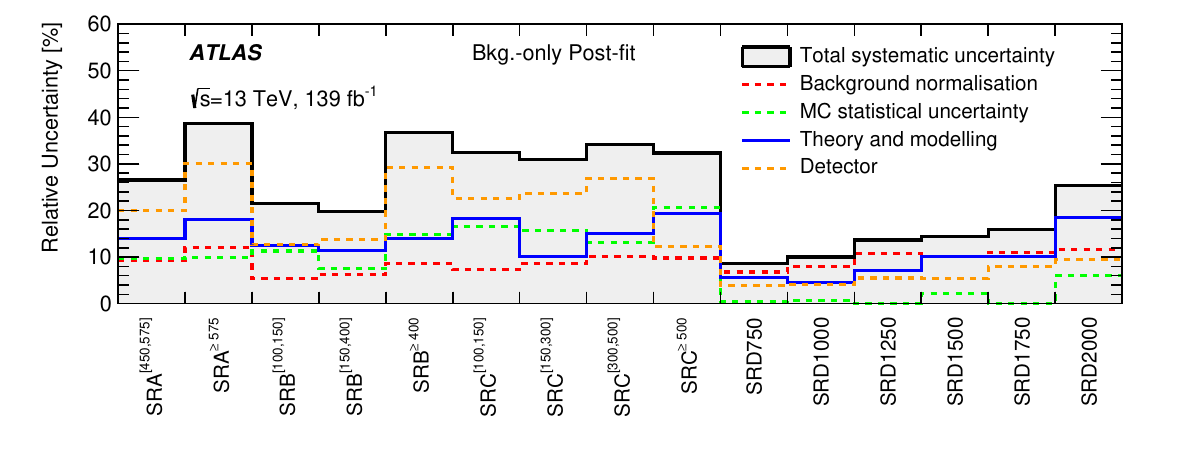}
\end{center}
\caption{Summary of the systematic uncertainties affecting the background yields in the SRs after the likelihood fit to data (`Post-fit'). The ``Detector'' category contains all detector-related systematic uncertainties.  The ``Background normalisation'' represents the uncertainty in the fitted normalisation factors, including the available data event counts in the CRs. The ``Theory and modelling'' represents the theoretical uncertainties of the modelling of the SM background processes. The individual uncertainties may not sum (in quadrature) to the total uncertainty due to correlations between the different components. Each superscript stands for the limits of each bin defined for the corresponding SR as presented in Section~\ref{sec:analysis}. For instance, SRA$^{[450,575]}$ refers to events in the SRA region with \mttwobinc\ $\in [450,575] \GeV$. }
\label{fig:SystematicPlot}
\end{figure}

The breakdown of the systematic uncertainties in the post-fit background prediction is shown in Figure \ref{fig:SystematicPlot}. The contributions are split by model-dependent SR. The total uncertainty shown is not simply the sum in quadrature of the individual uncertainties due to correlations between the components resulting from the fit. In the A-type regions, the dominant uncertainty arises from the experimental uncertainties of the large-$R$ jets, driven by the relatively high-$p_{\mathrm{T}}$ large-$R$ jets in this region of parameter space. In the B-type and C-type regions, the main contribution is from the small-$R$ jet uncertainties due to the tight selections on the leading $b$- and leading $c$-jet $p_{\mathrm{T}}$. In the D-type regions, the main uncertainty is the uncertainty in the background normalisation parameters, as, in comparison to the other SRs, the CRDs and SRD are kinematically much closer. This generally results in a reduction of the impact of the uncertainties in the SRs, absorbed by the normalization parameters of the background, leaving the uncertainty from the background normalisation in the CRs as the dominant systematic uncertainty. The statistical uncertainties on the MC also contribute to the total uncertainty of the background, particularly in SRA, SRB and SRC due to the tight selection criteria of these regions.
 
\FloatBarrier
 
\section{Results}

The presence of supersymmetric signals is explored by performing multi-bin fits that maximize a likelihood function, $\mathcal{L}(\mu,\boldsymbol{\theta})$, constructed as a product of Poisson probabilities for all bins considered in the search. This likelihood depends on a parameter of interest $\mu_{\mathrm{sig}} = \sigma^\text{fit}/\sigma^\text{theory}$, defined as the ratio of the fitted signal cross-section being tested ($\sigma^\text{fit}$) and its theoretical prediction ($\sigma^\text{theory}$). The effects of many sources of systematic uncertainty in both the signal and background yields are included when performing the likelihood fit through the introduction of nuisance parameters that impact the expectation values of the poissonian terms for each CR and SR bin. The nuisance parameters are described by Gaussian probability density functions, with the standard deviations on the functions corresponding to a specific experimental or theoretical modelling uncertainty. The preferred value of each nuisance parameter is determined as a part of the likelihood fit. Unconstrained normalization parameters ($\mu_{\mathrm{bkg}}$) are also defined to adjust the background predictions in the kinematic regime probed by this search. The fits performed do not significantly alter or constrain the nuisance parameter values relative to the fit input.
 
Three likelihood fits are performed: the ``background-only'' fit, which corrects the SM background prediction in all regions of the analysis by performing a fit only in the CRs and applying the resulting normalisation parameters to the SM prediction in the VRs and SRs; the ``model-dependent'' fit, which uses both of the CRs and SRs to evaluate the confidence-levels (CLs) for a specific signal hypothesis; and the ``model-independent'' fit, which is used to calculate the $p$-value of the SM-only hypothesis~\cite{Baak:2014wma}, again, using both the CRs and the SRs.
 
\begin{figure}[htpb]
\begin{center}
\includegraphics[width=0.9\textwidth]{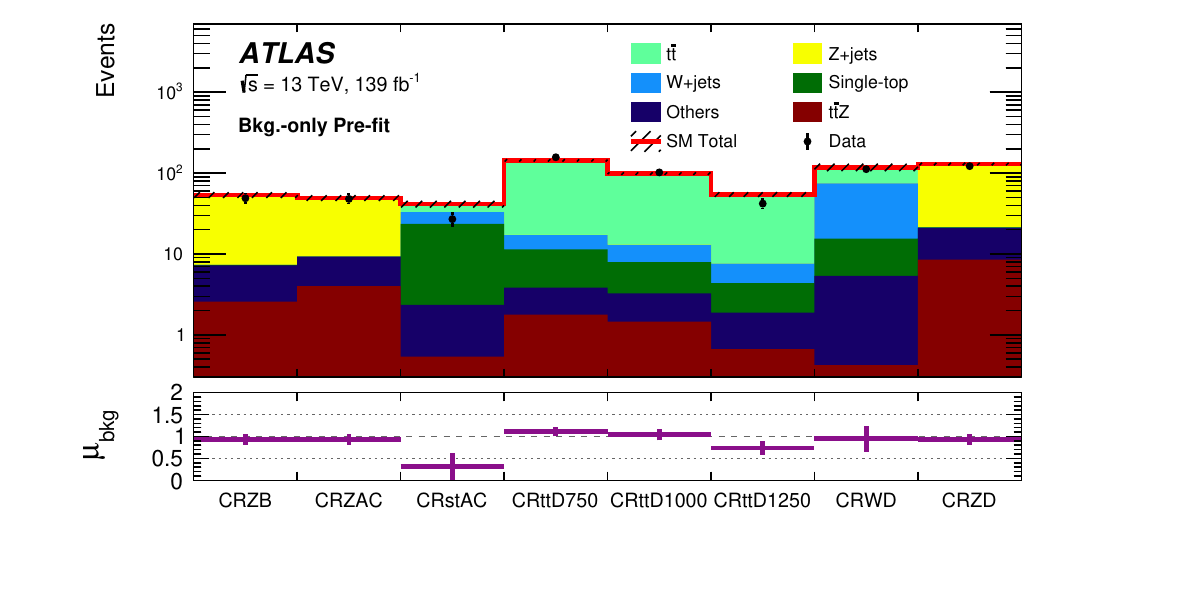}
\end{center}
\caption{Control region data and SM prediction yields before performing the background-only fit (top-panel) and the obtained normalisations after performing the background-only fit (bottom panel). Generally, the normalisation parameters are found to be consistent with unity. Both systematic and statistical uncertainties are considered in the uncertainty band. The $\mu_{\mathrm{bkg}}$ label represents the normalisation parameter for the SM background calculated from the relevant CR, for example in CRZB, $\mu_{\mathrm{bkg}}$ represents the normalisation for the \Zjets background.}
\label{fig:CRABC_Pulls}
\end{figure}

When performing the background-only fit, the CRs are used in the likelihood, and the fitted background estimate is then compared with the observed yields in the VRs and SRs. A single fit is performed using the associated CRs for the A, B, C, and D regions. The SM processes which do not have an associated CR are derived from the MC prediction, but they are allowed to vary within their own uncertainties in the fit. Concerning the A, B, and C regions, there are two unconstrained normalisation parameters, for the $\Zjets$ and single-top backgrounds. For the D regions, there are five unconstrained normalisation parameters, one for \Wjets, one for \Zjets, and three normalisations (split over the different \meff bins corresponding to the SRs) for the \ttbar background. Due to the different phase-space targeted by the SRD region, the normalisation factors calculated from the ABC regions are applied solely to the ABC regions, and similarly for the D regions. The top panel of Figure~\ref{fig:CRABC_Pulls} presents the pre-fit agreement between the data and the SM predictions in the CRs, which is subsequently used to estimate the background normalization parameters shown in the bottom panel. These normalisations are mostly consistent with unity, aside from the single-top normalisation. A relatively large uncertainty is also found for the single-top normalisation, driven by the comparison between the DR and DS calculation schemes. As the single-top background is sub-dominant and is a relatively small contribution to the SRs in comparison to the \Zjets background, the single-top normalisation and associated uncertainty is not of great concern.
 
\begin{figure}[htpb]
\begin{center}
\includegraphics[width=0.9\textwidth]{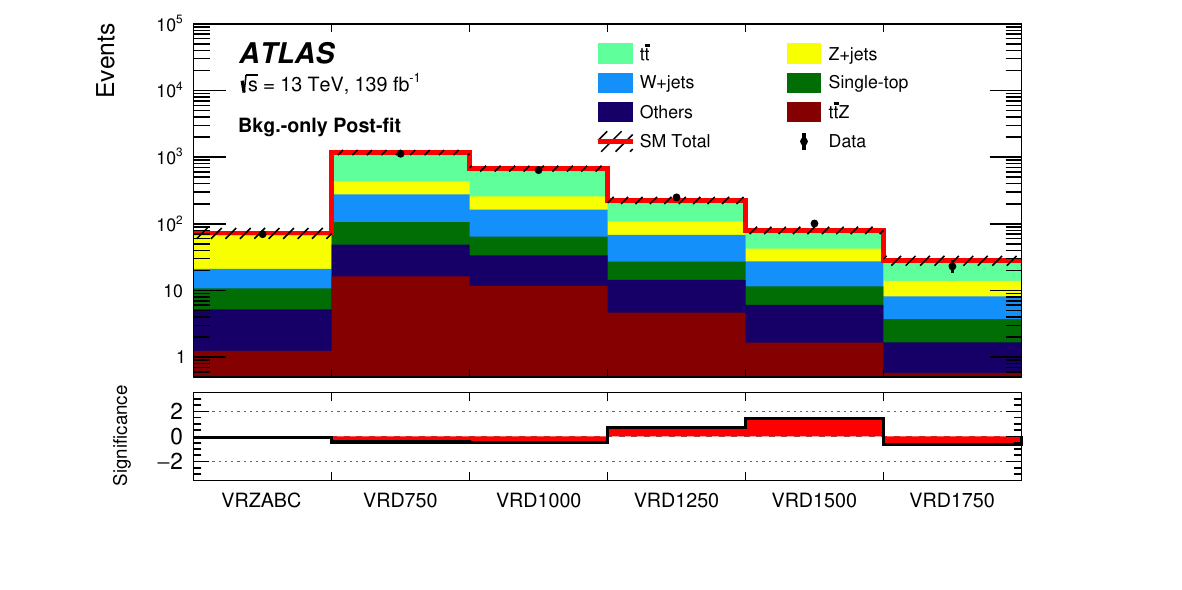}
\end{center}
\caption{Validation region yields and statistical significance \cite{Cousins:2007bmb}, derived from the background-only fit. The post-fit VR yields are found to be consistent within 2$\sigma$ of the observed data. Both systematic and statistical uncertainties are considered in the uncertainty band.}
\label{fig:VRABC_Pulls}
\end{figure}

\begin{figure}[htpb]
\begin{center}
\includegraphics[width=0.9\textwidth]{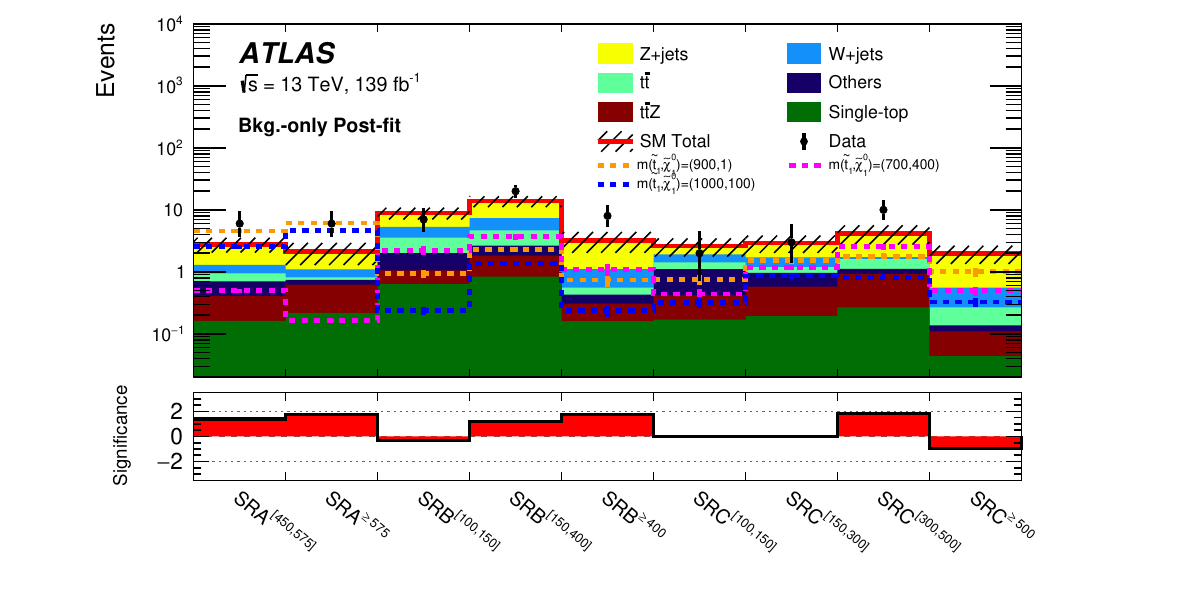}
\end{center}
\caption{Background-only post-fit SM yields and statistical significance in the model-dependent SRA, SRB, and SRC selections. The largest deviation between the post-fit expectation and the observed data is close to 2$\sigma$ in three regions. Both systematic and statistical uncertainties are considered in the uncertainty band. The pre-fit contribution for representative signal scenarios in each analysis region is shown for illustrative purposes.}
\label{fig:SRAB_Pulls}
\end{figure}

\begin{figure}[htpb]
\begin{center}
\includegraphics[width=0.9\textwidth]{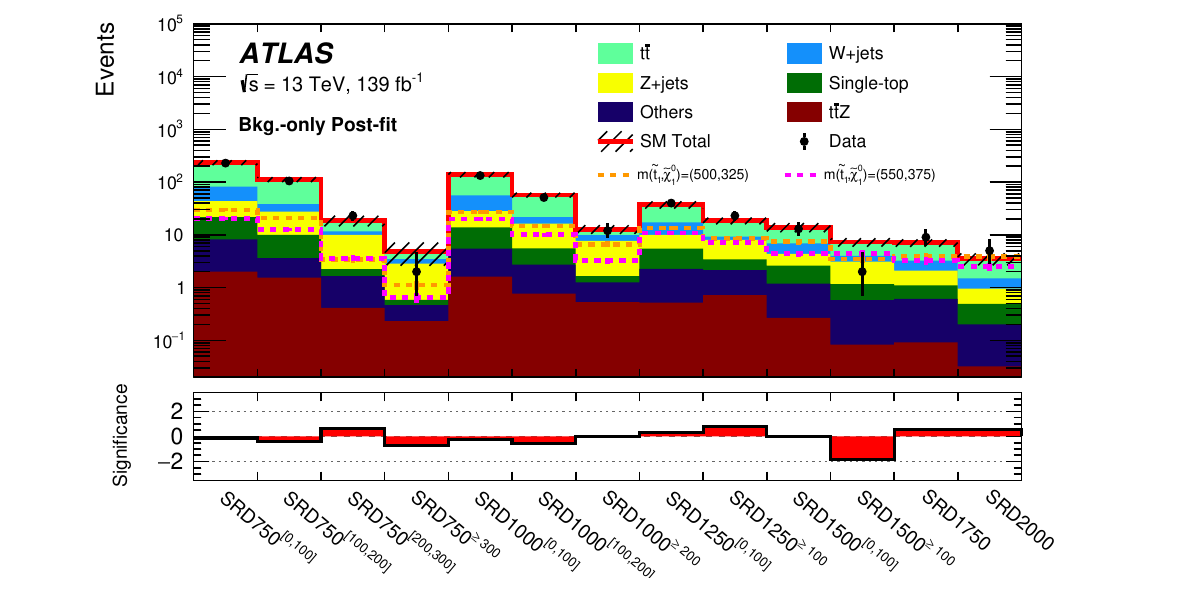}
\end{center}
\caption{Background-only post-fit SM yields and statistical significance in the model-dependent SRD selections. The largest deviation between the post-fit expectation and the observed data is a 1.8$\sigma$ deficit in the SRD1500 with \mtjclose $\geq 100 $ \GeV. Both systematic and statistical uncertainties are considered in the uncertainty band. The pre-fit contribution for representative signal scenarios in each analysis region is shown for illustrative purposes.}
\label{fig:SRC_Pulls}
\end{figure}

Generally, good agreement is observed between data and post-fit background in all VRs, which is shown in Figure~\ref{fig:VRABC_Pulls}. The statistical significance~\cite{Cousins:2007bmb} of the deviations between the observed data and the post-fit SM is also evaluated in all VRs, showing a maximum deviation of less than 2$\sigma$, confirming the good modelling of the main background provided by the fit strategy. Finally, Figures~\ref{fig:SRAB_Pulls} and ~\ref{fig:SRC_Pulls} present the post-fit SM yields and observed data in the SRs using the selections and the bins for the model-dependent fit defined in Section~\ref{sec:analysis}. The largest background contribution in the SRA and SRB regions is \Zjets, followed by single-top. There are deviations from the SM prediction in the fit to the SRs, with a largest deviation of $2 \sigma$, generally corresponding to the SRs which contain the tightest \mttwobinc selections. The dominant background in the SRD regions is \ttbar, followed by \Vjets and then single-top production. The post-fit SM is in very good agreement with the data with a largest deficit of $1.8\sigma$ in the SRD1500 high-\mtjclose bin.

\begin{figure}[htpb]
\centering
\subfloat[][]{
\includegraphics[width=0.47\textwidth]{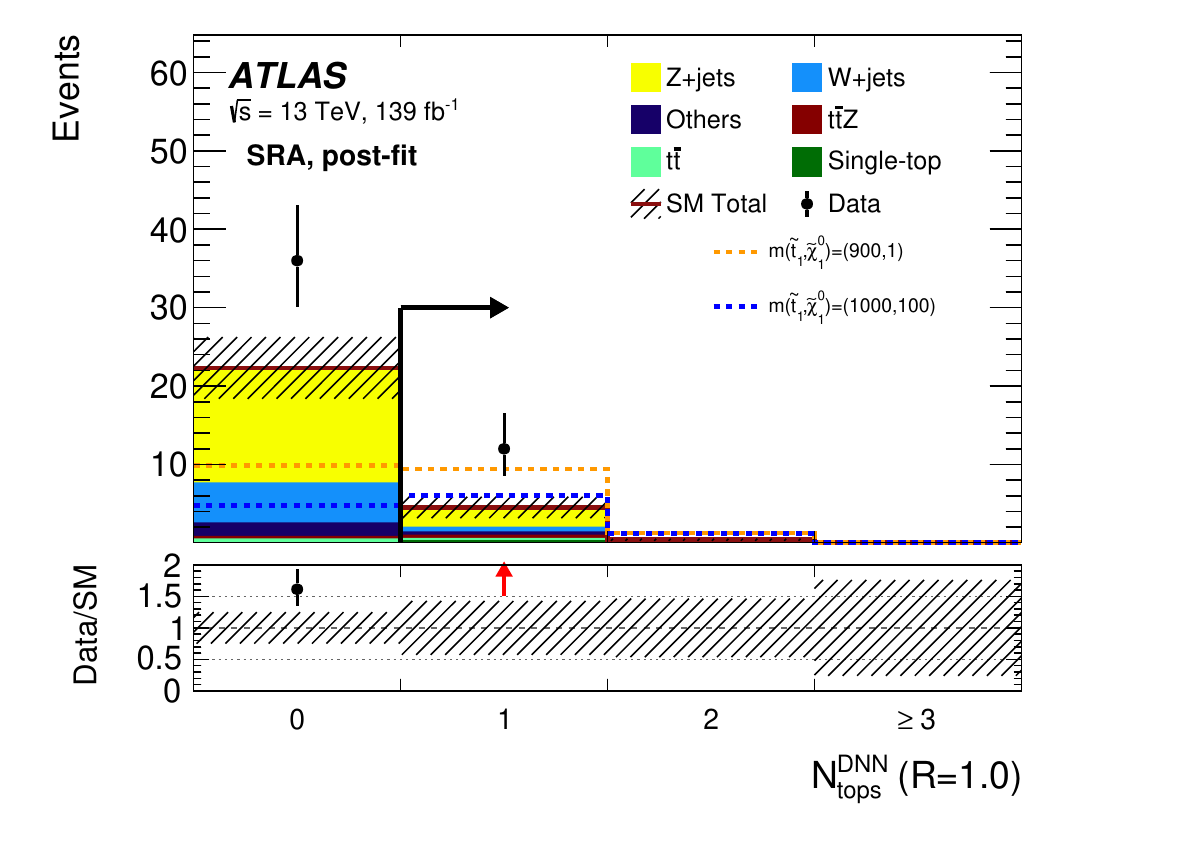}
\label{subfig:sra_ntop}
}
\subfloat[][]{
\includegraphics[width=0.47\textwidth]{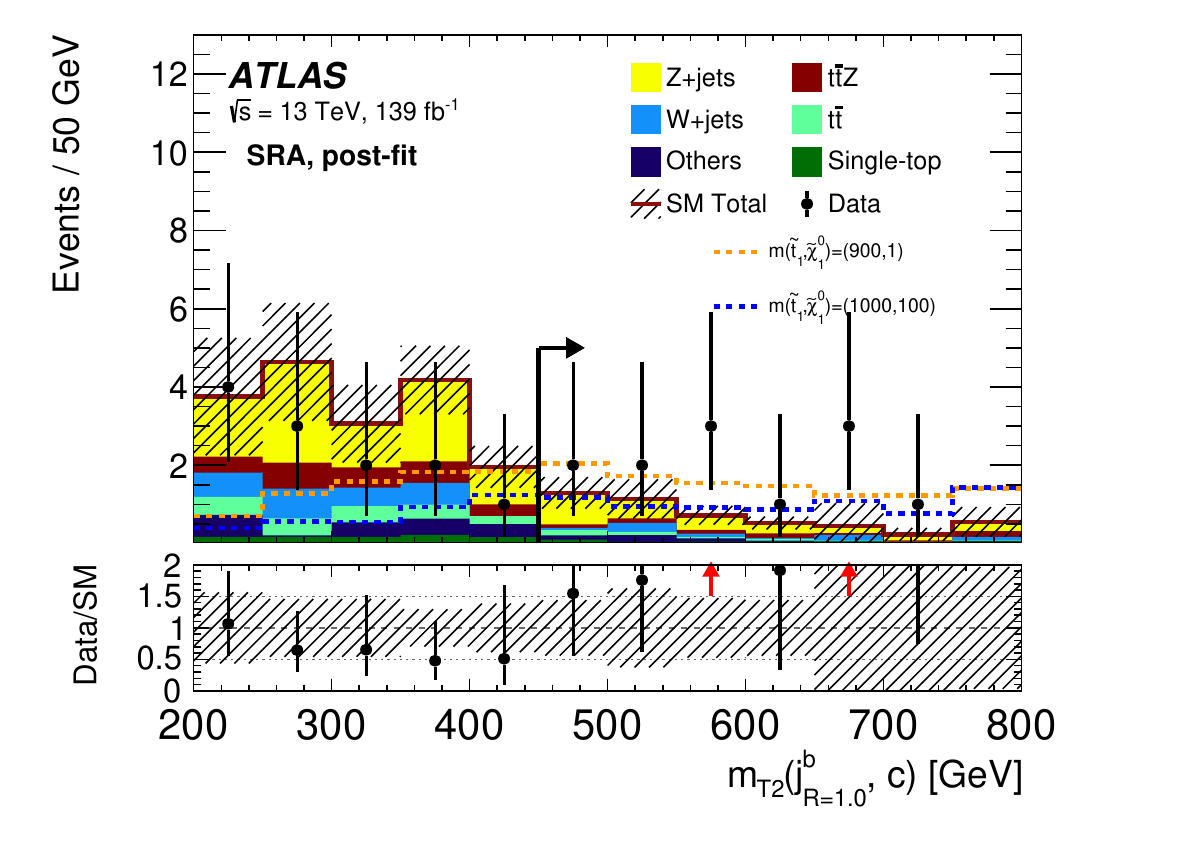}
\label{subfig:sra_mttwo}
} \\
\subfloat[][]{
\includegraphics[width=0.47\textwidth]{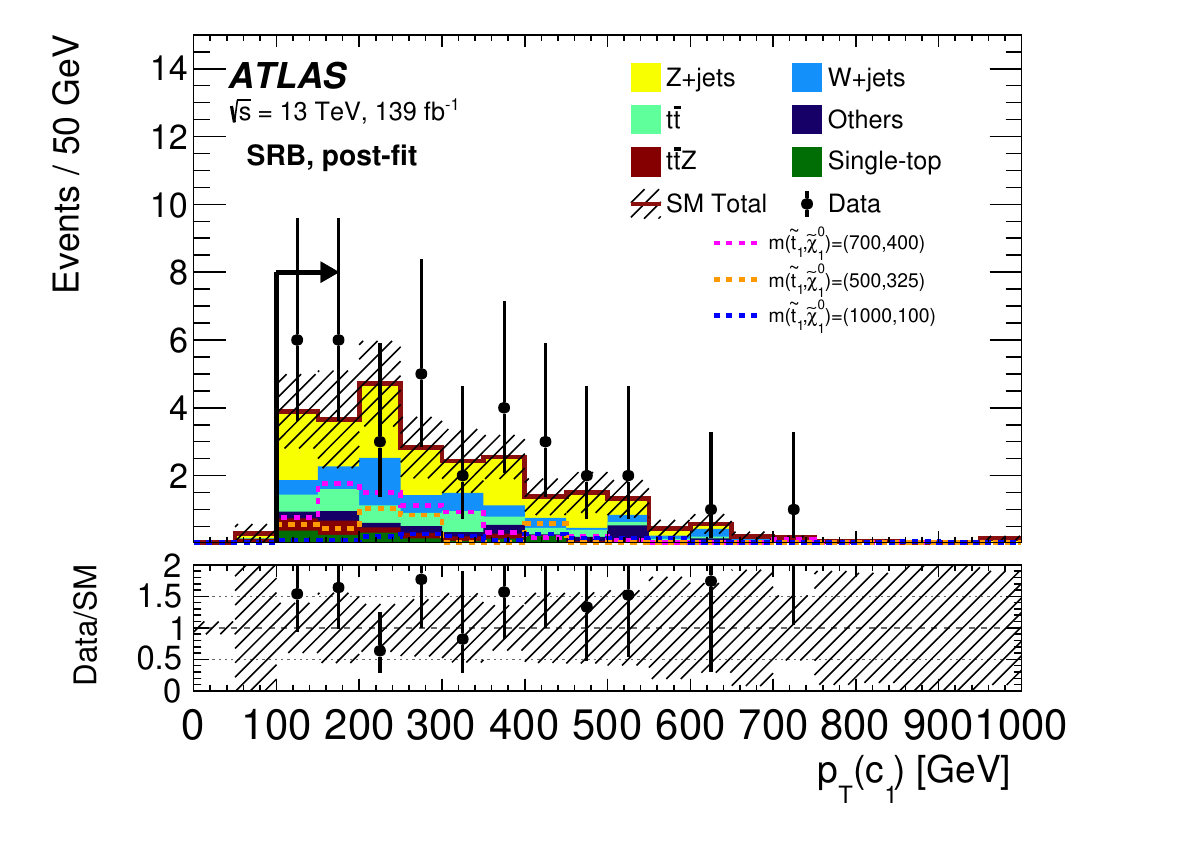}
\label{subfig:srb_ptc}
}
\subfloat[][]{
\includegraphics[width=0.47\textwidth]{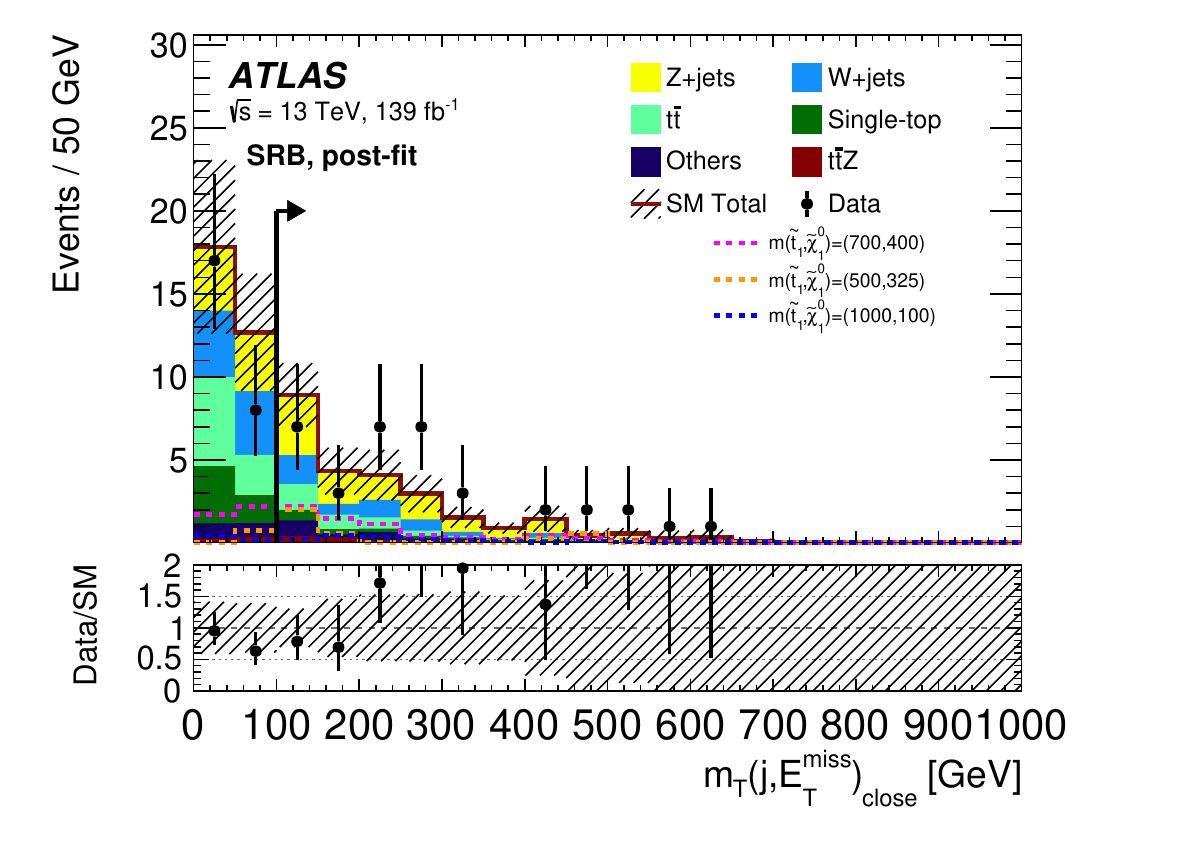}
\label{subfig:srb_mtj}
}\\
\caption{A selection of kinematic distributions in SRA and SRB, presented without the associated SR selection applied to the variable under consideration: \protect\subref{subfig:sra_ntop} presents $N_{\mathrm{tops}}^{\mathrm{DNN}}$ in SRA; \protect\subref{subfig:sra_mttwo} \mttwobinc in SRA; \protect\subref{subfig:srb_ptc} \ptc{1} in SRB; \protect\subref{subfig:srb_mtj} \mtjclose in SRB.  The selection applied on the given variable is represented by the arrow. The right-most bin in each histogram contains the overflow entries. The expected distributions for representative signal scenarios in each analysis region are shown for illustrative purposes. }
\label{fig:kinematics_sr_one}
\end{figure}

\begin{figure}[htpb]
\begin{center}
 
\subfloat[][]{
\includegraphics[width=0.47\textwidth]{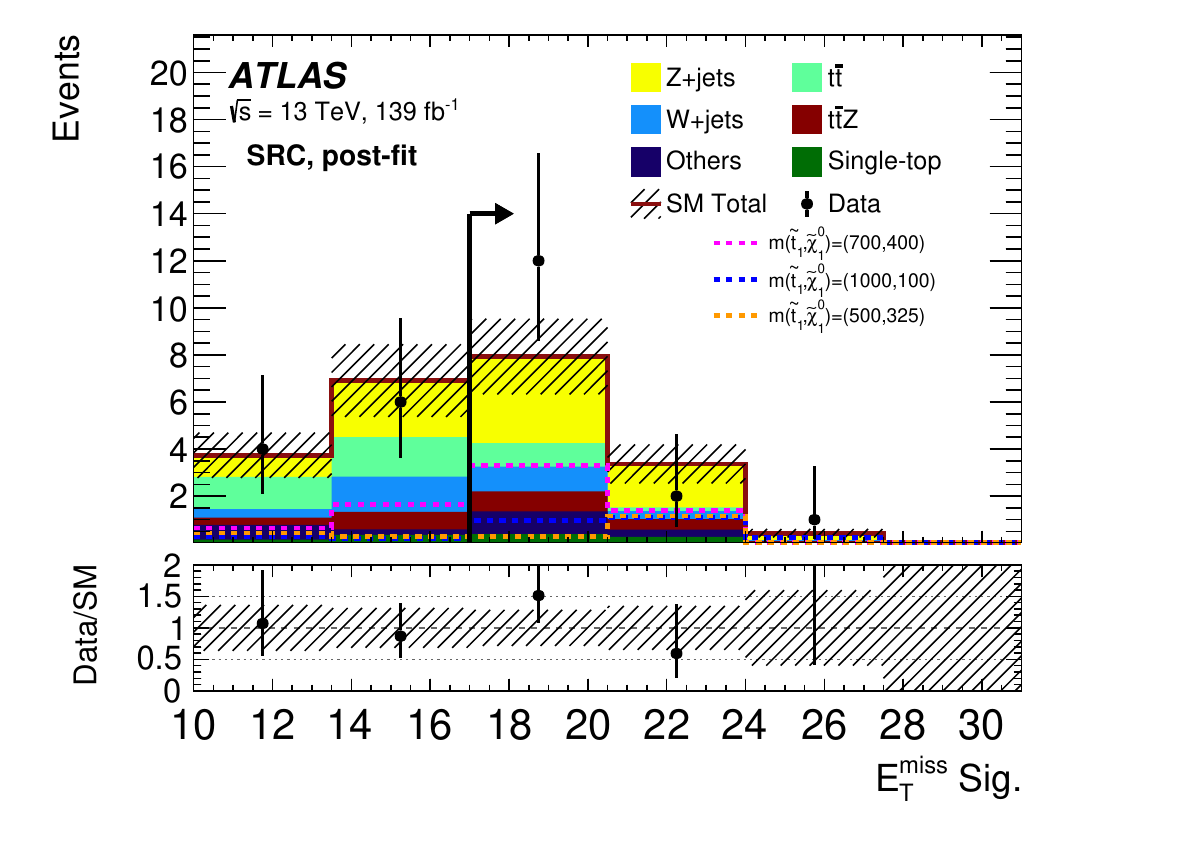}
\label{subfig:src_metsig}
}
\subfloat[][]{
\includegraphics[width=0.47\textwidth]{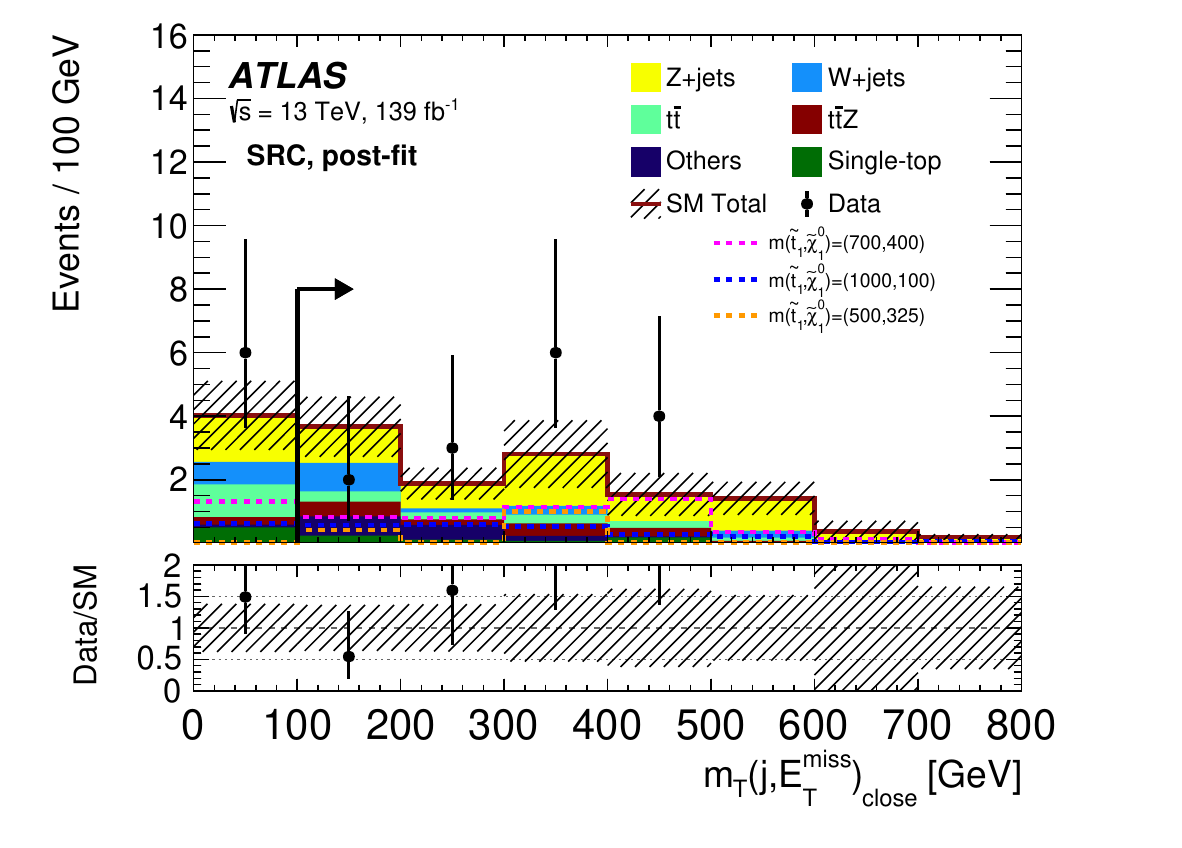}
\label{subfig:src_mtj}
}\\
\subfloat[][]{
\includegraphics[width=0.47\textwidth]{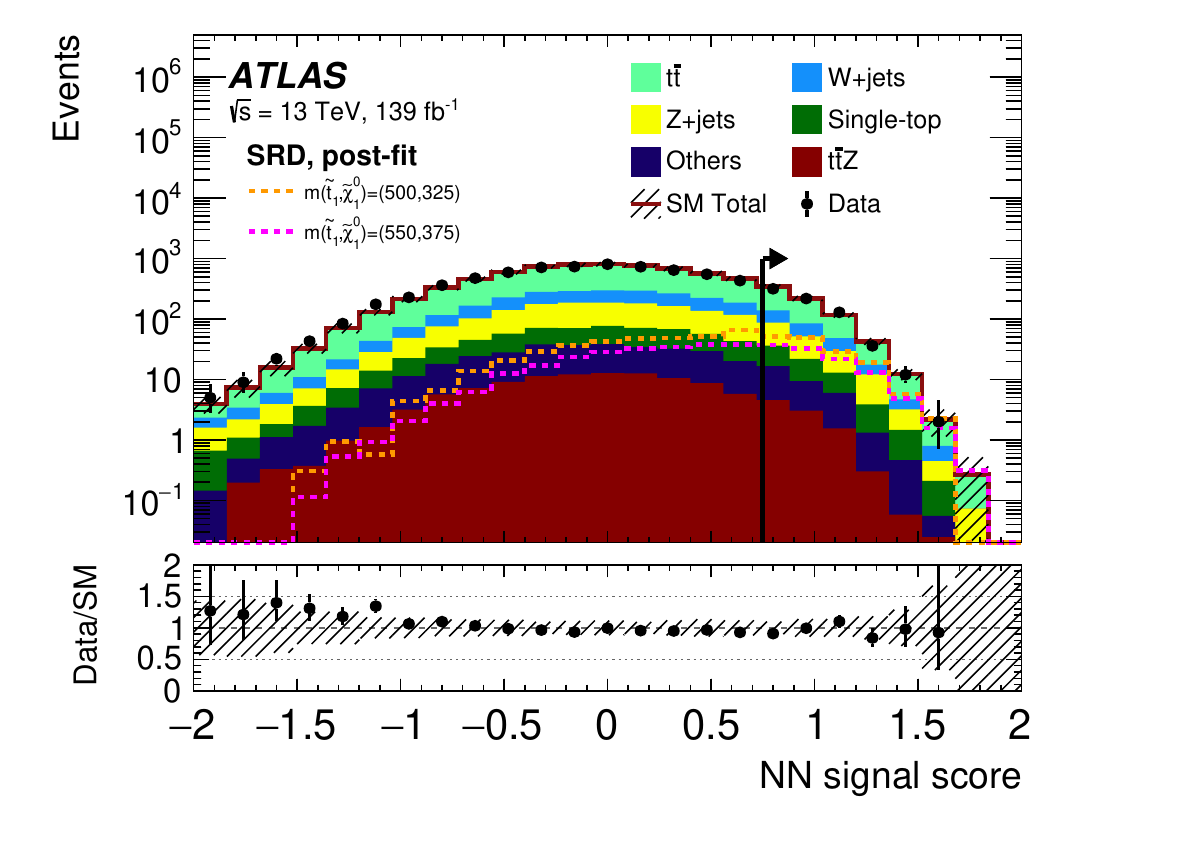}
\label{subfig:srd_nn}
}
\subfloat[][]{
\includegraphics[width=0.47\textwidth]{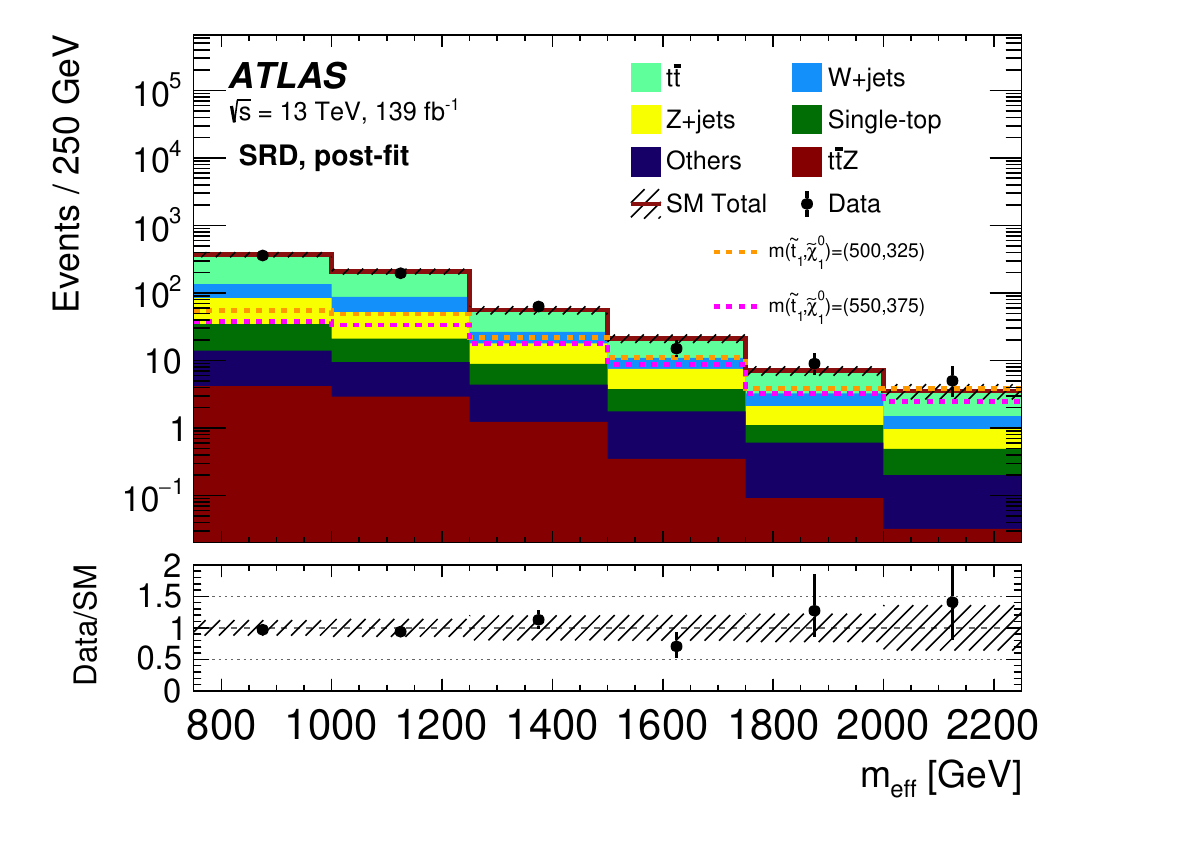}
\label{subfig:srd_meff}
}\\
\end{center}
\caption{A selection of kinematic distributions in SRC and SRD, presented without the associated SR selection applied to the variable under consideration (except for the SRD \meff distribution at the bottom right plot): \protect\subref{subfig:src_metsig} $\met$ Sig in SRC; \protect\subref{subfig:src_mtj} \mtjclose in SRC; \protect\subref{subfig:srd_nn} NN signal score in SRD; \protect\subref{subfig:srd_meff} \meff in SRD. The selection applied on the given variable is represented by the arrow. The right-most bin in each histogram contains the overflow entries. The expected distributions for representative signal scenarios in each analysis region are shown for illustrative purposes. }
\label{fig:kinematics_sr_two}
\end{figure}

Figures~\ref{fig:kinematics_sr_one} and ~\ref{fig:kinematics_sr_two} present a selection of key kinematic variables in the SRs, where the arrow shown denotes where the selection on that variable is applied in the relevant region. The $\sim 2 \sigma$ discrepancies in the SRA and SRC regions are already present in these distributions. For the SRB and SRD selections, it is seen that generally there is good agreement between the post-fit SM prediction and the observed data.
 
 
\begin{figure}[htpb]
\begin{center}
\includegraphics[width=1.0\textwidth]{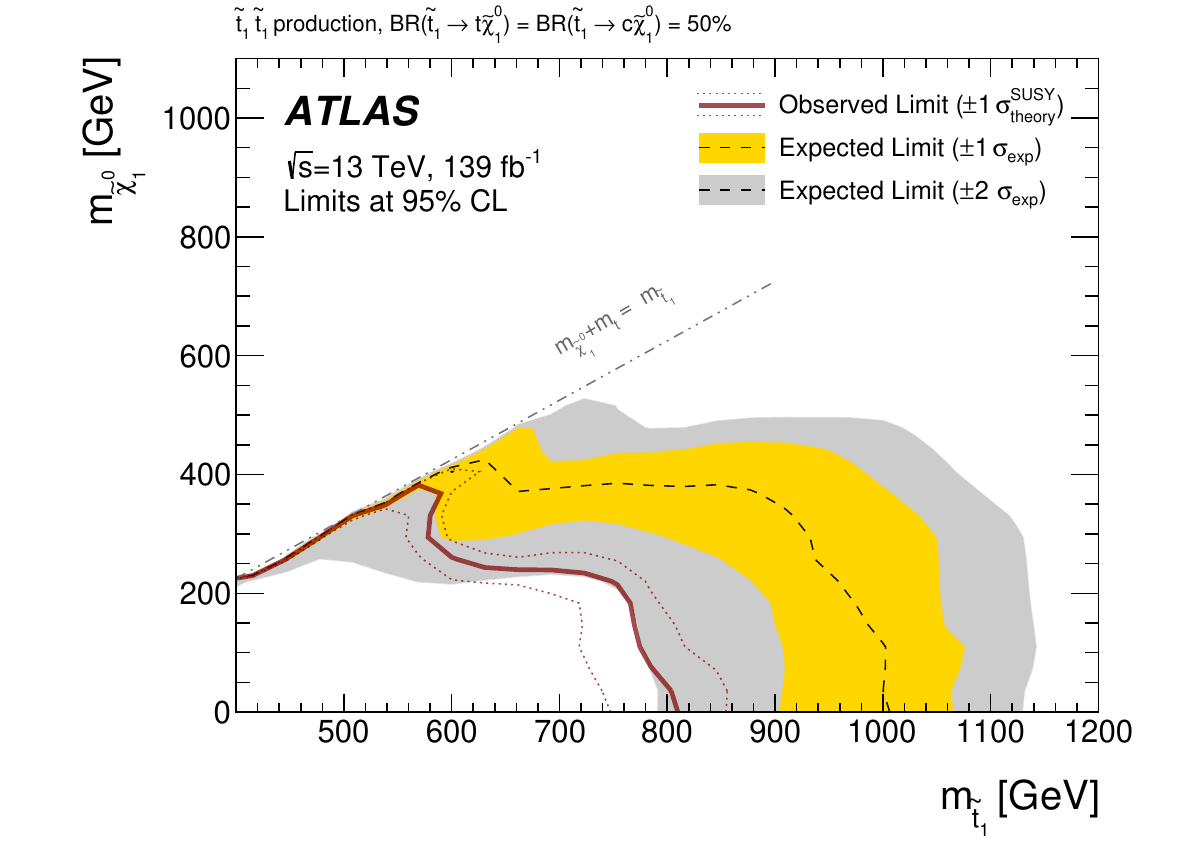}
\end{center}
\caption{Exclusion limits at the 95\% CL in the $m(\stopone)$ -- $m(\ninoone)$ plane, assuming BR$(\stopone \rightarrow t + \ninoone) = 0.5$. The dashed line, yellow band, and grey band present the expected limit, $\pm1\sigma$ uncertainty, and $\pm2\sigma$ uncertainty respectively. The solid red line presents the observed upper limit on the signal cross-section. The dashed red lines present the observed upper limit if the signal cross-section is varied by $\pm1\sigma$ of its predicted theoretical uncertainty.}
\label{fig:ExclusionLimitMass}
\end{figure}

\begin{figure}[htpb]
\begin{center}
\includegraphics[width=1.0\textwidth]{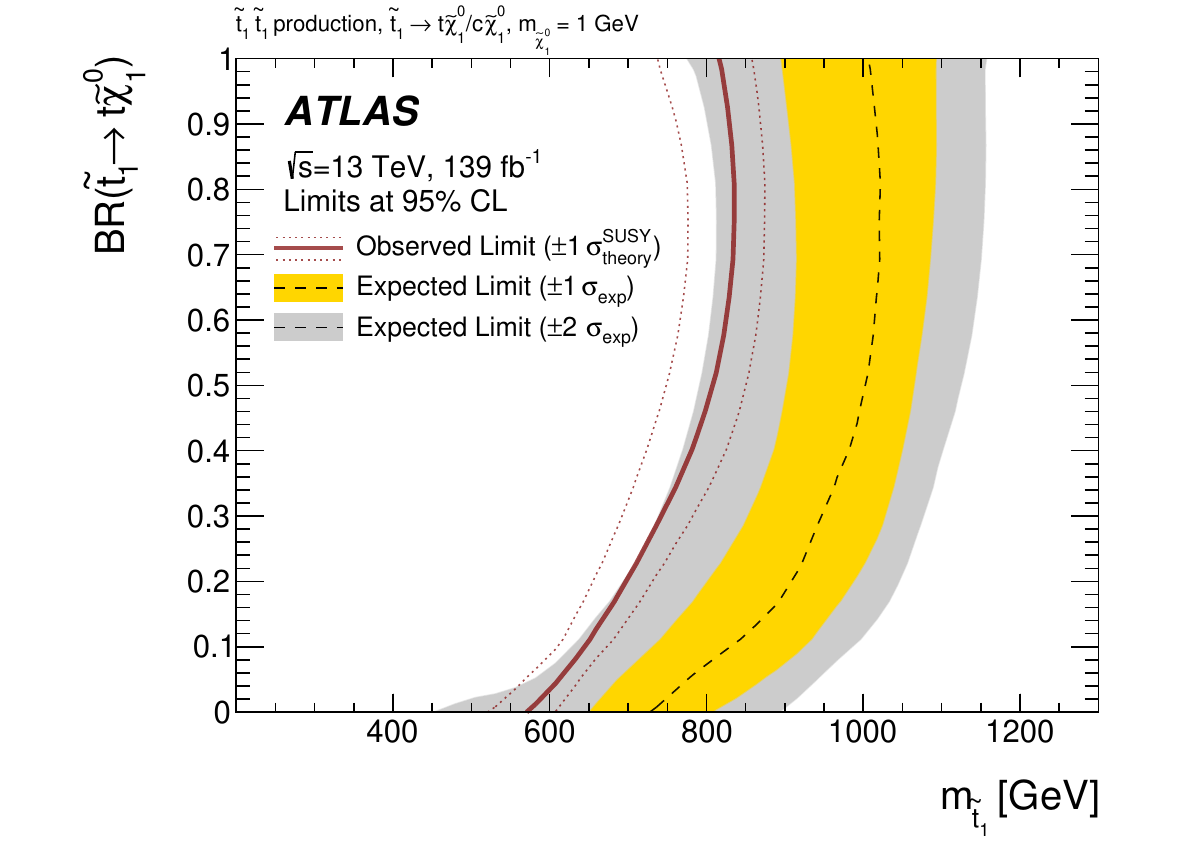}
\end{center}
\caption{Exclusion limits at the 95\% CL in the $m(\stopone)$, BR$(\stopone \rightarrow t + \ninoone)$ plane, assuming $m(\ninoone) = 1$~\GeV. The dashed line, yellow band, and grey band present the expected limit, $\pm1\sigma$ uncertainty, and $\pm2\sigma$ uncertainty respectively. The solid red line presents the observed upper limit on the signal cross-section. The dashed red lines present the observed upper limit if the signal cross-section is varied by $\pm1\sigma$ of its predicted theoretical uncertainty.}
\label{fig:ExclusionLimitBR}
\end{figure}

The model-dependent fit takes into account the contribution of the specific signal model that is being considered in all CRs and SRs to derive 95\% CL exclusion limits on the SUSY signal scenarios. These limits are obtained by performing a combined fit to all SRs, i.e. SRA, SRB, SRC and SRD, increasing the sensitivity to the SUSY signal models considered by using the orthogonal bins as described in Section~\ref{sec:analysis}. The result of this combined fit is presented in Figure ~\ref{fig:ExclusionLimitMass}, where the BR($\stopone \rightarrow t + \ninoone$)$=0.5$ is chosen, as this scenario gives the maximal number of $tc+\met$ events. The effect of the $1.8\sigma$ over-fluctuation of data in the A and C regions is clearly observed. Despite this, an exclusion up to 800\,\GeV on top-squark masses is observed for a massless neutralino. In the compressed region, top-squark masses up to 600\,\GeV are excluded. Figure~\ref{fig:ExclusionLimitBR} presents an alternative interpretation, where the neutralino mass is fixed to 1\,\GeV and the BR of the $\stopone \rightarrow t + \ninoone$ decay is allowed to vary between zero and one. As expected, the maximal sensitivity in this interpretation is obtained for BR($\stopone \rightarrow t + \ninoone$)$= 0.5$. A relatively high sensitivity is still found as the BR moves to 1 for the $\stopone \rightarrow t + \ninoone$ decay, primarily arising from the identification of $c$-jets from $W$ boson decays. In this scenario, for BR$(\stopone \rightarrow t + \ninoone) \geq 0.5$, top-squark masses are excluded up to 800~\GeV, whereas in the case where BR$(\stopone \rightarrow t + \ninoone) = 0$, the mass exclusion is reduced to 600~\GeV.
 
\pagebreak
The model-independent fit is performed by including the signal regions with an inclusive binning to maximise the general sensitivity to new phenomena, as previously discussed. A profile-likelihood-ratio statistic is used with a signal strength of $\mu_{\mathrm{sig}}$ assumed to contribute only to the SR, and is used to assess the $p$-value of the background-only hypothesis, and to extract observed and expected 95\% confidence level (CL) limits on the number of signal events ($S_{\mathrm{obs}}^{95}$ and $S_{\mathrm{exp}}^{95}$ respectively). The 95\% CL limit on the observed visible cross-section ($\epsilon\sigma^{95}_{\mathrm{obs}}$, where $\epsilon$ denotes the efficiency times acceptance) is calculated by dividing $S_{\mathrm{obs}}^{95}$ by the total integrated luminosity. Table~\ref{tab:discoveryFitResults} presents these results for the inclusive SRs. The calculated $p$-values (and thus model-independent limits) generally reflect the $1.8\sigma$ differences between the observed and expected yields in the SRs.

\renewcommand{\arraystretch}{1.5}
\begin{table}
\centering
\setlength{\tabcolsep}{0.0pc}
\caption[Breakdown of upper limits.]{95\% CL upper limits on the visible cross-section ($\langle\epsilon\sigma\rangle_{\mathrm{obs}}^{95}$) and on the number of signal events ($S_{\mathrm{obs}}^{95}$ ).  The third column ($S_{\mathrm{exp}}^{95}$) shows the 95\% CL upper limit on the number of signal events, given the expected number (and $\pm 1\sigma$ shifts of the expectation) of background events. The last two columns indicate the $CL_{\mathrm{B}}$ value, which provides a measure of compatibility of the observed data with the 95\% CL signal strength hypothesis relative to the fluctuations of the background, and the discovery $p$-value ($p_0$) together with is corresponding Gaussian significance ($Z$), which measures compatibility of the observed data with the background-only (zero signal strength) hypothesis relative to fluctuations of the background.}
\begin{tabular*}{\textwidth}{@{\extracolsep{\fill}}lccccc}
\noalign{\smallskip}\hline\noalign{\smallskip}
\textbf{Signal region}                        & $\langle\epsilon{\mathrm \sigma}\rangle_{\mathrm{obs}}^{95}$[fb]  &  $S_{\mathrm{obs}}^{95}$  & $S_{\mathrm{exp}}^{95}$ & $CL_{\mathrm{B}}$ & $p_0$ ($Z$)  \\
\noalign{\smallskip}\hline\noalign{\smallskip}
\noalign{\smallskip}\hline\noalign{\smallskip}
SRA (\mttwobinc $\geq 450$ \GeV)    & $0.10$ &  $14.4$ & $ { 8.4 }^{ +3.5 }_{ -1.9 }$ & $0.94$ & $ 0.02$~$(2.1)$ \\%
SRA (\mttwobinc $\geq 575$ \GeV)    & $0.07$ &  $9.4$ & $ { 5.8 }^{ +2.9 }_{ -1.2 }$ & $0.89$ & $ 0.04$~$(1.7)$ \\%
SRB (\mtjclose $\geq 100$ \GeV)    & $0.17$ &  $24.1$ & $ { 16.8 }^{ +7.0 }_{ -5.2 }$ & $0.85$ & $ 0.09$~$(1.3)$ \\%
SRB (\mtjclose $\geq 150$ \GeV)    & $0.16$ &  $22.8$ & $ { 13.2 }^{ +5.5 }_{ -3.6 }$ & $0.95$ & $ 0.03$~$(1.9)$ \\%
SRB (\mtjclose $\geq 400$ \GeV)    & $0.08$ &  $11.3$ & $ { 6.5 }^{ +3.1 }_{ -1.6 }$ & $0.92$ & $ 0.04$~$(1.8)$ \\%
SRC (\mtjclose $\geq 100$ \GeV)    & $0.09$ &  $12.6$ & $ { 9.6 }^{ +4.2 }_{ -2.1 }$ & $0.76$ & $ 0.22$~$(0.76)$ \\%
SRC (\mtjclose $\geq 150$ \GeV)    & $0.09$ &  $11.9$ & $ { 8.7 }^{ +3.9 }_{ -1.9 }$ & $0.81$ & $ 0.15$~$(1.0)$ \\%
SRC (\mtjclose $\geq 300$ \GeV)    & $0.08$ &  $11.0$ & $ { 7.8 }^{ +3.6 }_{ -1.7 }$ & $0.83$ & $ 0.13$~$(1.1)$ \\%
SRC (\mtjclose $\geq 500$ \GeV)    & $0.02$ &  $2.5$ & $ { 4.0 }^{ +2.4 }_{ -1.4 }$ & $0.13$ & $ 0.50$~$(0.00)$ \\%
SRD (\meff $\geq 750$ \GeV,~\mtjclose $\geq 200$ \GeV)    & $0.15$ &  $20.4$ & $ { 18.5 }^{ +8.4 }_{ -5.1 }$ & $0.58$ & $ 0.50$~$(0.00)$ \\%
SRD (\meff $\geq 1000$ \GeV,~\mtjclose $\geq 200$ \GeV)    & $0.10$ &  $13.9$ & $ { 13.7 }^{ +3.5 }_{ -5.7 }$ & $0.52$ & $ 0.50$~$(0.00)$ \\%
SRD (\meff $\geq 1250$ \GeV)   & $0.30$ &  $41$ & $ { 37 }^{ +12 }_{ -11 }$ & $0.60$ & $ 0.50$~$(0.00)$ \\%
SRD (\meff $\geq 1500$ \GeV)   & $0.09$ &  $12.9$ & $ { 14.6 }^{ +6.3 }_{ -4.1 }$ & $0.36$ & $ 0.50$~$(0.00)$ \\%
SRD (\meff $\geq 1750$ \GeV)   & $0.09$ &  $12.1$ & $ { 9.1 }^{ +3.9 }_{ -1.9 }$ & $0.77$ & $ 0.20$~$(0.84)$ \\%
SRD (\meff $\geq 2000$ \GeV)   & $0.05$ &  $7.3$ & $ { 5.6 }^{ +3.0 }_{ -1.2 }$ & $0.70$ & $ 0.26$~$(0.64)$ \\%
\noalign{\smallskip}\hline\noalign{\smallskip}
\end{tabular*}
\label{tab:discoveryFitResults}
\end{table}
\renewcommand{\arraystretch}{1.}


\FloatBarrier
 
\section{Conclusion}

This paper presented a first search for \stopone pair production leading to signatures with mixed final state containing a top-quark, a charm-quark and missing transverse momentum. Signal regions are defined based on recent top- and charm-tagging techniques developed by the ATLAS Collaboration.  Neural networks are also used to increase the discrimination of backgrounds for compressed signal scenarios, traditionally difficult to probe. The SM background prediction is corrected in dedicated control regions and verified in validation regions, both  depleted of signal contributions but representative of the kinematics of the SRs.
 
A multi-bin profile likelihood fit is performed to assess the agreement of the signal and SM predictions against data. No significant deviations are observed from the expected background prediction. The largest deviation reaches a significance of $1.8\sigma$ in the SRs  targeting the bulk and intermediate region of parameter space. In the optimal scenario for this analysis, where the branching ratios for the $\stopone \rightarrow t + \ninoone$ and $\stopone \rightarrow c + \ninoone$ are equal, an exclusion on the top-squark masses up to 800\,\GeV\ is found for light neutralinos. In the compressed region, considering the same branching ratio scenario, top-squark masses up to 600\,\GeV\ are excluded. These constitute the first results to date at the LHC on a search for BSM physics in this final-state signature.


\section*{Acknowledgements}


We thank CERN for the very successful operation of the LHC and its injectors, as well as the support staff at
CERN and at our institutions worldwide without whom ATLAS could not be operated efficiently.
 
The crucial computing support from all WLCG partners is acknowledged gratefully, in particular from CERN, the ATLAS Tier-1 facilities at TRIUMF/SFU (Canada), NDGF (Denmark, Norway, Sweden), CC-IN2P3 (France), KIT/GridKA (Germany), INFN-CNAF (Italy), NL-T1 (Netherlands), PIC (Spain), RAL (UK) and BNL (USA), the Tier-2 facilities worldwide and large non-WLCG resource providers. Major contributors of computing resources are listed in Ref.~\cite{ATL-SOFT-PUB-2023-001}.
 
We gratefully acknowledge the support of ANPCyT, Argentina; YerPhI, Armenia; ARC, Australia; BMWFW and FWF, Austria; ANAS, Azerbaijan; CNPq and FAPESP, Brazil; NSERC, NRC and CFI, Canada; CERN; ANID, Chile; CAS, MOST and NSFC, China; Minciencias, Colombia; MEYS CR, Czech Republic; DNRF and DNSRC, Denmark; IN2P3-CNRS and CEA-DRF/IRFU, France; SRNSFG, Georgia; BMBF, HGF and MPG, Germany; GSRI, Greece; RGC and Hong Kong SAR, China; ISF and Benoziyo Center, Israel; INFN, Italy; MEXT and JSPS, Japan; CNRST, Morocco; NWO, Netherlands; RCN, Norway; MEiN, Poland; FCT, Portugal; MNE/IFA, Romania; MESTD, Serbia; MSSR, Slovakia; ARRS and MIZ\v{S}, Slovenia; DSI/NRF, South Africa; MICINN, Spain; SRC and Wallenberg Foundation, Sweden; SERI, SNSF and Cantons of Bern and Geneva, Switzerland; MOST, Taipei; TENMAK, T\"urkiye; STFC, United Kingdom; DOE and NSF, United States of America.
 
Individual groups and members have received support from BCKDF, CANARIE, CRC and DRAC, Canada; CERN-CZ, PRIMUS 21/SCI/017 and UNCE SCI/013, Czech Republic; COST, ERC, ERDF, Horizon 2020, ICSC-NextGenerationEU and Marie Sk{\l}odowska-Curie Actions, European Union; Investissements d'Avenir Labex, Investissements d'Avenir Idex and ANR, France; DFG and AvH Foundation, Germany; Herakleitos, Thales and Aristeia programmes co-financed by EU-ESF and the Greek NSRF, Greece; BSF-NSF and MINERVA, Israel; Norwegian Financial Mechanism 2014-2021, Norway; NCN and NAWA, Poland; La Caixa Banking Foundation, CERCA Programme Generalitat de Catalunya and PROMETEO and GenT Programmes Generalitat Valenciana, Spain; G\"{o}ran Gustafssons Stiftelse, Sweden; The Royal Society and Leverhulme Trust, United Kingdom.
 
In addition, individual members wish to acknowledge support from Chile: Agencia Nacional de Investigaci\'on y Desarrollo (FONDECYT 1190886, FONDECYT 1210400, FONDECYT 1230987); China: National Natural Science Foundation of China (NSFC - 12175119, NSFC 12275265, NSFC-12075060); Czech Republic: PRIMUS Research Programme (PRIMUS/21/SCI/017); European Union: European Research Council (ERC - 948254), Horizon 2020 Framework Programme (MUCCA - CHIST-ERA-19-XAI-00), Italian Center for High Performance Computing, Big Data and Quantum Computing (ICSC, NextGenerationEU), Marie Sklodowska-Curie Actions (EU H2020 MSC IF GRANT NO 101033496); France: Agence Nationale de la Recherche (ANR-20-CE31-0013, ANR-21-CE31-0013, ANR-21-CE31-0022), Investissements d'Avenir Idex (ANR-11-LABX-0012), Investissements d'Avenir Labex (ANR-11-LABX-0012); Germany: Baden-Württemberg Stiftung (BW Stiftung-Postdoc Eliteprogramme), Deutsche Forschungsgemeinschaft (DFG - CR 312/5-1); Italy: Istituto Nazionale di Fisica Nucleare (FELLINI G.A. n. 754496, ICSC, NextGenerationEU); Japan: Japan Society for the Promotion of Science (JSPS KAKENHI JP21H05085, JSPS KAKENHI JP22H01227, JSPS KAKENHI JP22H04944); Netherlands: Netherlands Organisation for Scientific Research (NWO Veni 2020 - VI.Veni.202.179); Norway: Research Council of Norway (RCN-314472); Poland: Polish National Agency for Academic Exchange (PPN/PPO/2020/1/00002/U/00001), Polish National Science Centre (NCN 2021/42/E/ST2/00350, NCN OPUS nr 2022/47/B/ST2/03059, NCN UMO-2019/34/E/ST2/00393, UMO-2020/37/B/ST2/01043, UMO-2021/40/C/ST2/00187); Slovenia: Slovenian Research Agency (ARIS grant J1-3010); Spain: BBVA Foundation (LEO22-1-603), Generalitat Valenciana (Artemisa, FEDER, IDIFEDER/2018/048), La Caixa Banking Foundation (LCF/BQ/PI20/11760025), Ministry of Science and Innovation (MCIN \& NextGenEU -PCI2022-135018-2, MICIN \& FEDER -PID2021-125273NB, RYC2019-028510-I, RYC2020-030254-I, RYC2021-031273-I), PROMETEO and GenT Programmes Generalitat Valenciana (CIDEGENT/2019/023, CIDEGENT/2019/027); Sweden: Swedish Research Council (VR 2018-00482, VR 2022-03845, VR 2022-04683, VR grant 2021-03651), Knut and Alice Wallenberg Foundation (KAW 2017.0100, KAW 2018.0157, KAW 2018.0458, KAW 2019.0447); Switzerland: Swiss National Science Foundation (SNSF - PCEFP2\_194658); United Kingdom: Leverhulme Trust (Leverhulme Trust RPG-2020-004); United States of America: Neubauer Family Foundation.


\clearpage

\printbibliography
 
\clearpage
 
\begin{flushleft}
\hypersetup{urlcolor=black}
{\Large The ATLAS Collaboration}

\bigskip

\AtlasOrcid[0000-0002-6665-4934]{G.~Aad}$^\textrm{\scriptsize 102}$,
\AtlasOrcid[0000-0002-5888-2734]{B.~Abbott}$^\textrm{\scriptsize 120}$,
\AtlasOrcid[0000-0002-1002-1652]{K.~Abeling}$^\textrm{\scriptsize 55}$,
\AtlasOrcid[0000-0001-5763-2760]{N.J.~Abicht}$^\textrm{\scriptsize 49}$,
\AtlasOrcid[0000-0002-8496-9294]{S.H.~Abidi}$^\textrm{\scriptsize 29}$,
\AtlasOrcid[0000-0002-9987-2292]{A.~Aboulhorma}$^\textrm{\scriptsize 35e}$,
\AtlasOrcid[0000-0001-5329-6640]{H.~Abramowicz}$^\textrm{\scriptsize 151}$,
\AtlasOrcid[0000-0002-1599-2896]{H.~Abreu}$^\textrm{\scriptsize 150}$,
\AtlasOrcid[0000-0003-0403-3697]{Y.~Abulaiti}$^\textrm{\scriptsize 117}$,
\AtlasOrcid[0000-0002-8588-9157]{B.S.~Acharya}$^\textrm{\scriptsize 69a,69b,m}$
% ...
\bigskip
\\

% ...
$^{29}$Physics Department, Brookhaven National Laboratory, Upton NY; United States of America.\\
% ...
$^{35}$$^{(a)}$Facult\'e des Sciences Ain Chock, R\'eseau Universitaire de Physique des Hautes Energies - Universit\'e Hassan II, Casablanca;$^{(b)}$Facult\'{e} des Sciences, Universit\'{e} Ibn-Tofail, K\'{e}nitra;$^{(c)}$Facult\'e des Sciences Semlalia, Universit\'e Cadi Ayyad, LPHEA-Marrakech;$^{(d)}$LPMR, Facult\'e des Sciences, Universit\'e Mohamed Premier, Oujda;$^{(e)}$Facult\'e des sciences, Universit\'e Mohammed V, Rabat;$^{(f)}$Institute of Applied Physics, Mohammed VI Polytechnic University, Ben Guerir; Morocco.\\
% ...
$^{49}$Fakult\"{a}t Physik , Technische Universit{\"a}t Dortmund, Dortmund; Germany.\\
% ...
$^{55}$II. Physikalisches Institut, Georg-August-Universit\"{a}t G\"ottingen, G\"ottingen; Germany.\\
% ...
$^{69}$$^{(a)}$INFN Gruppo Collegato di Udine, Sezione di Trieste, Udine;$^{(b)}$ICTP, Trieste;$^{(c)}$Dipartimento Politecnico di Ingegneria e Architettura, Universit\`a di Udine, Udine; Italy.\\
% ...
$^{102}$CPPM, Aix-Marseille Universit\'e, CNRS/IN2P3, Marseille; France.\\
% ...
$^{117}$Department of Physics, New York University, New York NY; United States of America.\\
% ...
$^{120}$Homer L. Dodge Department of Physics and Astronomy, University of Oklahoma, Norman OK; United States of America.\\
% ...
$^{150}$Department of Physics, Technion, Israel Institute of Technology, Haifa; Israel.\\
$^{151}$Raymond and Beverly Sackler School of Physics and Astronomy, Tel Aviv University, Tel Aviv; Israel.\\
% ...
$^{a}$ Also Affiliated with an institute covered by a cooperation agreement with CERN.\\
$^{b}$ Also at An-Najah National University, Nablus; Palestine.\\
$^{c}$ Also at Borough of Manhattan Community College, City University of New York, New York NY; United States of America.\\
$^{d}$ Also at Center for High Energy Physics, Peking University; China.\\
$^{e}$ Also at Center for Interdisciplinary Research and Innovation (CIRI-AUTH), Thessaloniki; Greece.\\
$^{f}$ Also at Centro Studi e Ricerche Enrico Fermi; Italy.\\
% ...
$^{m}$ Also at Department of Physics, King's College London, London; United Kingdom.\\
% ...

\end{flushleft}


\end{document}